\documentclass[10pt]{article}
\usepackage{amsmath,amssymb,subfigure}
\usepackage[dvips]{epsfig}
   
\newtheorem{lem}{Lemma}[section]

\newtheorem{rem}[lem]{Remark}

\newtheorem{prop}[lem]{Proposition}
\newtheorem{theo}[lem]{Theorem}
\newcommand{\proof}{{\bf Proof}:\quad}
\newcommand{\nc}{\newcommand}
\newcommand{\rnc}{\renewcommand}
\nc{\beqn}{\begin{eqnarray*}}
\nc{\eeqn}{\end{eqnarray*}}
\nc{\be}{\begin{equation}}
\nc{\ee}{\end{equation}}
\nc{\beqa}{\begin{eqnarray}}
\nc{\eeqa}{\end{eqnarray}}
\rnc{\a}{\alpha}
\rnc{\b}{\beta}
\rnc{\d}{\delta}
\nc{\ga}{\gamma}
\nc{\lb}{\lambda}
\nc{\p}{\psi}
\nc{\e}{\epsilon}
\rnc{\c}{\chi}
\nc{\sg}{\sigma}
\rnc{\t}{\theta}
\nc{\om}{\omega}
\rnc{\P}{\Psi}
\nc{\G}{\Gamma}
\nc{\tw}{\tilde{w}}
\nc{\tP}{\tilde{P}^g}
\nc{\tQ}{\tilde{Q}^g}
\nc{\dpt}{d\tilde{p}^g(\lb)}
\nc{\dqt}{d\tilde{q}^g(\lb)}

\nc{\ra}{\rightarrow}
\nc{\Ra}{\Rightarrow}
\nc{\LRa}{\Leftrightarrow}
\nc{\lra}{\leftrightarrow}
\nc{\ot}{\otimes}

\nc{\mat}[4]{\left(\begin{array}{cc}#1&#2\\#3&#4\end{array}\right)}
\nc{\ca}{\mathcal{A}}
\nc{\cc}{\mathcal{C}}
\nc{\f}{\mathcal{F}}
\nc{\cn}{\mathcal{N}}
\nc{\cs}{\mathcal{S}}
\nc{\g}{\mathcal{G}}
\nc{\Rc}{\mathcal{R}}
\nc{\gtp}{\tilde{\mathcal{G}}^{\prime}(\lb)}
\nc{\Gt}{\tilde{\Gamma}}
\nc{\yl}{y(\lb)}
\nc{\yz}{y^+(z)}
\newcommand{\dozl}{\omega^g_z(\lb)}

\nc{\csh}{\hat{\cs}_g}
\nc{\pl}{\sqrt{R(\lb)}}
\nc{\rhoap}{(2\rho^{\prime}(z)-2\a^{\prime}(z))}
\nc{\py}{\sqrt{R(y)}}
\nc{\px}{\sqrt{R(\xi)}}
\nc{\dolz}{\omega^g_{\lb}(z)}
\nc{\dorz}{\omega^g_{\lb}(z)}
\nc{\dozr}{\omega^g_{z}(\lb)}
\nc{\dhorz}{\hat{\omega}^g_r(z)}
\nc{\Up}{\Upsilon}
\nc{\up}{\upsilon}

\newcommand{\Om}{\Omega}
\newcommand{\pie}{\dfrac{1}{2\pi i}}

\newcommand{\li}{\mathcal{L}}
\newcommand{\resi}{\operatornamewithlimits{Res}_{z=\infty}}
\newcommand{\vu}{{\vec{u}}}
\nc{\VU}{{\vec{U}}}
\newcommand{\Rn}{{\rm I\!R}}

\newcommand{\Pn}{{\rm I\!P}}

\begin{document}
\title{Riemann-Hilbert problem for  the small dispersion limit of the  KdV
equation and  linear overdetermined systems of Euler-Poisson-Darboux type}
\author{Tamara Grava\thanks{e--mail: t.grava@ic.ac.uk}\\
Department  of Mathematics\\
Imperial College of Science Technology and Medicine\\
 London, SW7 2BZ UK} 
\maketitle
\begin{abstract}
\noindent
We study the Cauchy problem for
 the Korteweg de Vries  (KdV) equation with small dispersion 
 and with monotonically increasing initial data
using the Riemann-Hilbert (RH) approach.  
The solution of the Cauchy problem, in the zero dispersion limit,
is obtained  using  the steepest descent  method for oscillatory 
Riemann-Hilbert problems.
The asymptotic solution is completely described by 
a scalar function $\g$ that satisfies a scalar RH problem and a set of
algebraic equations constrained by algebraic inequalities.
The scalar function $\g$ is equivalent to the solution of the 
 Lax-Levermore maximization  problem.
  The solution of the set
of algebraic equations satisfies
 the Whitham  equations.
We show that the scalar function $\g$
 and the Lax-Levermore maximizer  can be expressed as the 
solution of a linear overdetermined system of equations 
of Euler-Poisson-Darboux
 type.
We also show that the set of algebraic equations and algebraic
 inequalities can be expressed
in terms of solutions of  a different set of linear 
overdetermined systems
 of equations  of Euler-Poisson-Darboux  type. Furthermore
we show that the set of algebraic equations is equivalent to
the classical solution of the Whitham equations
expressed by the hodograph transformation.
\end{abstract}

\section{Introduction}
The Cauchy problem for the Korteweg de Vries
(KdV) equation  
\begin{equation}
\label{KdV}
u_t-6uu_x+\e^2u_{xxx}=0,\quad u(x,0)=u_0(x), \quad \e>0
\end{equation}
in the zero-dispersion limit has been widely studied. 
The physical interest in this limit is due to the fact 
that it describes the phenomenon
of  shock waves in dissipationsless dispersive media.
Dispersive shock waves are characterized by the appearance of rapid 
modulated  oscillations. Gurevich and Pitevskii \cite{GP} suggested that
  these oscillations could be modeled by  the solution of the
 one-phase Whitham equations \cite{W}.
The multiphase or $g$-phase   Whitham equations were derived by Flaschka,
Forest and Mc Laughlin \cite{FFM}.  Lax and Levermore \cite{LL}
rigorously showed that the multiphase Whitham equations appear
in the zero dispersion limit of the Cauchy problem for the KdV
equation with asymptotically reflectionless initial data.
Later   Venakides \cite{V} considered a wider class of initial data.
 Lax and Levermore  developed their theory  
in the frame of the zero-dispersion asymptotics  for the solution of the
inverse scattering problem  of KdV. They showed that 
the principal term of the relevant asymptotics 
  is given by the $g$-phase solution \cite{DMN} of the KdV equation 
 with the wave parameters depending on the 
functions $u_1(x,t)>\dots>u_{2g+1}(x,t)$ which satisfy the 
$g$-phase Whitham equations:
\begin{equation}
\dfrac{\partial u_i}{\partial t}-v_i(u_1,u_2,\dots,u_{2g+1})\dfrac{
\partial u_i}{\partial x}=0,\quad
x,t,u_i\in\Rn,\;\;\;i=1,...,2g+1,\;g\geq 0.
\label{whithamg}
\end{equation}
For $g>0$ the speeds  $v_i(u_1,u_2,\dots,u_{2g+1})$,
 $i=1,2,\dots,2g+1$,
depend  through $u_1,\dots,u_{2g+1}$ on complete hyperelliptic integrals
 of genus $g$. For $g=0$ we define $u_1=u$ and 
 the zero-phase Whitham equation  reads 
\begin{equation}
\label{zeroph}
\dfrac{\partial u}{\partial t}-6u\dfrac{\partial u}{\partial x}=0\,
\end{equation}
and can be integrated by the method of characteristics.
The formal integrability of equations (\ref{whithamg}) for $g>0$ 
 was obtained by
Tsarev \cite{T} 
using the geometric-Hamiltonian structure \cite{DN} of the Whitham
equations.   
Namely he proved that if
the functions  $w_i=w_i(u_1,u_2,\dots,u_{2g+1})$, $i=1,\dots,2g+1$,  
solve the linear over-determined system of equations
\begin{equation}
\label{tsarev0}
\dfrac{\partial w_i}{\partial u_j}=\dfrac{1}{v_i-v_j}\dfrac{\partial v_i}{\partial u_j}[w_i-w_j],\quad
i,j=1,2,\dots,2g+1,\;\;i\neq j,
\end{equation}
where $v_i=v_i(u_1,u_2,\dots,u_{2g+1}),\; i=1,\dots, 2g+1$, are 
the speeds in  (\ref{whithamg}), then 
 the solution $\vu(x,t)=(u_1(x,t),u_2(x,t),\dots,u_{2g+1}(x,t))$ 
of the so called  hodograph transformation
\begin{equation}
\label{ch00}
x=-v_i(\vu)\,t+w_i(\vu)
\,\quad i=1,\dots ,2g+1\,,
\end{equation}
satisfies equations (\ref{whithamg}). Conversely, any solution
$(u_1(x,t),u_2(x,t),\dots, u_{2g+1}(x,t))$  of (\ref{whithamg}) can be obtained in this
way in the neighborhood of $(x_0,t_0)$ at which $u_{ix}$'s are not
vanishing.
The general solution of the Tsarev equations  was  obtained 
 in \cite{FRT},\cite{G1},\cite{FRT1} for monotonically increasing initial data.
 The key step introduced in \cite{FRT},\cite{FRT1} was to
reduce
the solution of the Tsarev system to the solution of
linear-overdetermined 
systems of Euler-Poisson Darboux type for some functions $q_k=q_k(\vu)$,
 $\,k=1,\dots,g$, namely
\[
\dfrac{\partial}{\partial u_i}q_k- \dfrac{\partial}{\partial u_j}q_k=
2(u_i-u_j)\dfrac{\partial^2}{\partial u_i\partial u_j}q_k, \quad 
i\neq i,\; i,j=1,\dots,2g+1.
\]
In this paper  we show that these functions play a role in the
Lax-Levermore maximization problem.
We study the small dispersion limit of KdV for monotonically
increasing analytic 
initial data bounded at infinity.

\noindent
We use a different
 approach to  the small dispersion limit 
obtained in  \cite{DVZ} by Deift, Venakides and Zhou. 
These authors first  used the formulation of  the 
 Cauchy problem for KdV as a Riemann-Hilbert (RH) problem  \cite{S}
(see also \cite{BDT}). Then, for computing the small $\e$-asymptotics, 
they used the steepest descent  method for oscillatory Riemann-Hilbert problems
introduced in \cite{DZ}. 
 Their procedure
leads to a scalar RH problem for a certain phase function $\g$ that turns
out to be equivalent to  the solution of  the leading order variational problem
in the Lax-Levermore theory.  Furthermore they reduce the initial
value problem for the Whitham  equations to solving a set
of algebraic equations constrained by algebraic inequalities.
Existence and uniqueness of the initial value problem for the Whitham  
equations follows from the existence and uniqueness of the solution of the
variational problem  in the Lax-Levermore theory.

\noindent
The RH problem for the scalar function $\g$ is well
defined even  for smooth initial data.

\noindent
We show that, for smooth monotonically  increasing initial data
bounded at infinity,
the function $\g$ and  the Lax-Levermore maximizer can 
 be expressed as the  solution of a linear overdetermined system
of Euler-Poisson-Darboux type.
In the same way,  the set of algebraic equations obtained
through the Deift, Venakides and Zhou approach can be expressed as the
   solution of a set of  linear overdetermined systems of equations of
Euler-Poisson Darboux type.
 We also show that this set of algebraic equations is equivalent
to the set of algebraic equations defined by 
the hodograph transformation (\ref{ch00}).

\noindent
The advantage of this representation is clear when one tries to
construct effectively the solution of the set of algebraic equations
constrained by the algebraic inequalities. It is  simpler to
evaluate space derivatives and to estimate the sign of the quantities in the
inequalities.  Indeed this representation
was used in \cite{G1},\cite{G} to give an upper bound to the
genus of the solution of the Whitham equations.

This paper is organized as follows.

\noindent
Section 2 contains the definitions of the  Abelian differentials
on Riemann surface and the meromorphic analogue of the Cauchy
kernel.

\noindent
In Section 3  we review  the Riemann-Hilbert steepest descent method for the
zero dispersion KdV equation with monotonically increasing initial data.

\noindent
We determine the scalar function $\g$ associated to the RH problem
and all its properties in section 4. 
We show that the set of algebraic equations
that we call moment conditions and normalization conditions are
equivalent to the set of algebraic equations defined by the 
hodograph transformation.

\noindent
In Section 5 we show that the hodograph transformation obtained in Sec 4 is
equivalent to the classical one provided in \cite{FRT} and
\cite{G1}. 
 We then  show that the  moment conditions, the  normalization
conditions and the Lax-Levermore maximizer 
can be expressed in terms of solutions of linear overdetermined systems of
Euler-Poisson-Darboux
type. We also derive in a simple way the equations which determine the
phase transitions.

\noindent
Finally in section 6 we summarize the main results and draw our conclusions.

\section{Riemann surfaces and Abelian differentials: notations and
 definitions}
\setcounter{equation}{0}
Let
\begin{equation}
\cs_g:=\left\{P=(\lb,y),\;y^2=\prod_{j=1}^{2g+1}(\lb-u_j)\right\}\,,
\quad u_1>u_2>\dots>u_{2g+1}\,,\;\;u_i\in\Rn\,,
\end{equation}
be the hyperelliptic Riemann surface of genus $g\geq 0$. We shall use the 
standard
representation of $\cs_g$ as a two-sheeted covering of $C\Pn^1$ with  cuts
along the intervals
\begin{equation}
\label{cut}
[u_{2k},u_{2k-1}],\quad k=1,\dots,g+1,\quad u_{2g+2}=-\infty\,.
\end{equation}

\noindent
We choose the  basis $\{\alpha_j,\beta_j\}_{j=1}^g$ of 
the homology group $H_1(\cs_g)$ so that $\alpha_j$ lies fully on the
  upper sheet
and encircles clockwise the interval $[u_{2j},u_{2j-1}]$, $j=1,\dots,g$,
 while $\beta_j$ emerges on the upper sheet on the cut  $[u_{2j},u_{2j-1}]$,
  passes anti-clockwise to the lower sheet trough the cut $(-\infty, u_{2g+1}]$
 and returns to the initial point  through the lower sheet.

\noindent
The one-forms that are analytic on the closed Riemann surface $\cs_g$
except  for a finite number of points  are called Abelian differentials.

\noindent
We define on $\cs_g$ the following differentials \cite{S}:

\noindent
1) The canonical basis of holomorphic  one-forms or Abelian differentials
 of the first kind $\phi_1,\phi_2\dots\phi_g$:
\begin{equation}
\label{holo}
\phi_k(\lb)=\dfrac{\lb^{g-1}\gamma^k_1+\lb^{g-2}\ga^k_2+
\dots+\ga^k_g}{y(\lb)}d\lb\,,\quad k=1,\dots,g\,.
\end{equation}
The constants $\ga^k_i$ are uniquely determined by the 
 normalization  conditions  
\noindent
\begin{equation}
\label{snorm}
\int_{\alpha_j}\phi_k=\delta_{jk}\,,\quad j,k=1,\dots, g.
\end{equation}
We remark that an holomorphic differential having all its
$\alpha$-periods equal to zero is identically zero \cite{S}.

\noindent
2) The set $\sigma^g_k$, $k\geq 0$, $g\geq 0$, of Abelian differentials of the
second kind with a pole of order $2k+2$ at infinity, with asymptotic
behavior 
\begin{equation}
\label{sigma}
\sigma^g_k(\lb)=\left[\lb^{k-\frac{1}{2}}+O(\lb^{-\frac{3}{2}})\right]d\lb \quad 
\mbox{for
large}\;\; \lb
\end{equation}
  and normalized by the condition
\begin{equation}
\label{norm2}
\int_{\alpha_j}\sg^g_k=0,\quad j=1,\dots, g\,.
\end{equation}
We use the notation 
\begin{equation}
\label{dp}
\sg^g_0(\lb)=dp^g(\lb)\,,\quad\ 12\sg_1(\lb)=dq^g(\lb)\,\quad g\geq 0.
\end{equation} 
In literature the differentials $dp^g(\lb)$ and $dq^g(\lb)$ are called
quasi-momentum and quasi-energy respectively \cite{DN}.
The explicit formula for the differentials  $\sg^g_k$, $k\geq 0$, is given 
 by the expression
\beqa
\label{b6}
\sg^g_k(\lb)=\dfrac{P^g_k( \lb)}{y(\lb)}d\lb\;,
\quad P^g_k( \lb)= \lb^{g+k}+a^k_{1}
 \lb^{g+k-1}+a^k_{2} \lb^{g+k-2}\dots+a^k_{g+k}\,,
\eeqa
where  the coefficients 
$a^k_i=a^k_i(\vu)\;$,  $\vu=(u_1,u_2,\dots,u_{2g+1})$, $i=1,\dots,g+k,$ are uniquely determined by 
(\ref{sigma}) and (\ref{norm2}).

\noindent
3) The  Abelian differential of the
third kind  $\om_{qq_0}(\lb)$  with first order poles at the points 
$Q=(q,y(q))$ and $Q_0=(q_0,y(q_0))$  with residues
$\pm 1$ respectively. Its periods are normalized by the relation
\begin{equation}
\label{norm3}
\int_{\alpha_j}\om_{qq_0}(\lb)=0,\quad j=1,\dots,g\,.
\end{equation}
In the following we mainly use the normalized differential
 $\dozr$ which has  simple 
poles at the points $Q^{\pm}(z)=(z,\pm y(z))$
with residue $\pm 1$ respectively.

The differential  $\dozr$  is explicitly  given by the expression 
\begin{equation}
\label{CK}
\dozr=\dfrac{d\lb}{y(\lb)}\dfrac{y(z)}{\lb-z}-\sum_{k=1}^g\phi_k(\lb)
\int_{\alpha_k}\dfrac{dt}{y(t)}\dfrac{y(z)}{t-z}\,,
\end{equation}
where $\phi_k(\lb)$, $k=1,\dots,g$, 
is the  normalized basis of holomorphic differentials. By construction
\begin{equation}
\label{norm}
\int_{\a_j}\dozr=0,\quad j=1,\dots,g.
\end{equation}
The differential $\dozr$ can also be written in the form
\begin{equation}
\label{dob}
\dozr=\dfrac{d\lb}{y(\lb)}\dfrac{y(z)}{\lb-z}-\sum_{k=1}^gN^g_k(z,\vu)
\dfrac{\lb^{g-k}}
{y(\lb)}d\lb\,,
\end{equation}
where
\begin{equation}
\label{N}
N^g_k(z,\vu)=y(z)\sum_{j=1}^g\ga_{k}^j\int_{\a_j}\dfrac{d\eta}{y(\eta)(\eta-z)}.
\end{equation}
  $\dozr$  as a function of $z$, is an Abelian integral.
The periods of this integral are obtained from the relations \cite{Z}
\begin{equation}
\label{periods}
\int_{\alpha_j}d_z[\dozr]=0,\quad \int_{\beta_j}d_z[\dozr]=
4\pi i \phi_k(\lb)\,,\quad j=1,\dots g\,.
\end{equation}
The differential $\dozr$ satisfies the property \cite{Z}
\begin{equation}
\label{dozl}
d_z\dozl=d_{\lb}\om_{\lb}(z).
\end{equation}
In the following we will use the single value restriction of 
$\dozr$ determined by the conditions
\[
\dozr|_{z=u_{2g+1}}=0
\]
and by choosing  $y(z)$ to be analytic off the cuts (\ref{cut}) and
real positive  $z>u_{1}$.
We will still denote this single value restriction with
$\dozr$. We remark that $\dozr$ is a meromorphic analogue of the 
Cauchy kernel on the  Riemann surface $\cs_g$ \cite{Z}. 

The next proposition is also important for  our subsequent considerations.
\begin{prop}\cite{G}
\label{propog}
The Abelian  differentials of the second kind  $\sg^g_k(\lb)$, $\,k\geq
0$, defined in (\ref{sigma})
satisfy the relations
\begin{equation}
\label{maini}
\sg^g_k(\lb)=\dfrac{1}{2}\resi\left[\dozr\,z^{k-\frac{1}{2}}dz\right]=-\dfrac{1}{2k+1}d_{\lb}\resi
\left[ \dorz\,z^{k+\frac{1}{2}}\right]\,,
\end{equation}
where $\dozr$ has been defined in (\ref{CK}),
$\dorz$ is the normalized Abelian differential of the
 third kind with   simple
poles at the points $Q^{\pm}(\lb)=(\lb,\pm y(\lb))$
with residue $\pm 1$ respectively and $d_{\lb}$ denotes differentiation
with respect to $\lb$.
\end{prop}

%%%%%%%%%%%%%%%%%%%%%%%%%%%%%%%%%%%%%%%%%%%%%%%%%%%%%%%%%%%%%%%%%%%%%%%%%%%%%%%%%%%%%%%%%%%%%%%%%%%%%%%%%%%%%%%%%%%%%%%%%%%%%%%%%%%%%%%%%%%%%%%%%%%%%%%%%%%%%%%%%%%%%%%%%%%%%%%%%%%%%%%%%%%%%%%%%%%%%%%%%%%%%%%%%%%%%%%%%%%%%%%%%%%%%%%%%%%%%%%%%%%%%%%%%%%%%%%%%%%%%%%%%%%%%%%%%%%%%%%%%%%%%%%%%%%%%%%%%%%%%%%%%%%%%%%%%%%%%%%%%%%%%%%%%%%%%%%%%%%%%%%%%%%%%%%%%%%%%%%%%%%%%%%%%%%%%%%%%%%%%%%%%%%%
\section{Riemann-Hilbert steepest descent method for the
zero dispersion KdV equation with monotonically increasing initial data}
\setcounter{equation}{0}
Following \cite{S} we reformulate the inverse scattering for
the KdV equation as a RH problem.
 We consider
 monotonically increasing analytic initial data $u_0(x)$ bounded at
infinity. For convenience we assume
\[
\lim_{x\ra -\infty}u_0(x)=0, \quad \lim_{x\ra +\infty}u_0(x)=1.
\] 
We suppose
\begin{equation}
\label{const}
\int_{-\infty}^c u_0(x)(1+|x|^{2+\d})dx< \infty, \quad
\int_c^{-\infty}(1- u_0(x))(1+|x|^{2+\d})dx< \infty,
\end{equation}
for all finite $c$ and $\d>0$.

\noindent
Let  $r(\lb; \e), \;\lb>0$, be the reflection coefficient from the
left of the Sch\"odinger equation $-\e^2f_{xx}+u_0(x)f=\lb f$.
Define the matrix \cite{DVZ}, \cite{AC}
\beqa
\nonumber
\nu(\lb,\e)=\left\{
\begin{array}{lcl}
&\sg_1,& \quad \lb<0\\
&\begin{pmatrix}
0&-\bar{r}e^{-2i\alpha/\e}\\
r e^{2i\alpha/\e}&1\\
\end{pmatrix},& \quad 0<\lb<1,\\
&\begin{pmatrix}
1-|r|^2&-\bar{r}e^{-2i\alpha/\e}\\
r e^{2i\alpha/\e}&1\\
\end{pmatrix},& \quad \lb>1,
\end{array}\right.
\eeqa
where $\sg_1=\begin{pmatrix} 0&1\\1&0\\ \end{pmatrix}$ and $\alpha=
4t\lb^{\frac{3}{2}}+x\lb^{\frac{1}{2}}$. The goal is to find a row vector
valued function $m(\lb)=m(\lb;x,t,\e)=(m_1,m_2)$ analytic for complex
$\lb$ off the real axis, satisfying the jump and asymptotic conditions
\begin{equation}
\nonumber
\begin{split}
m_+(\lb;x,t,\e)=&m_-(\lb;x,t,\e)\nu(\lb;x,t,\e)\\
m(\lb;x,t,\e)\ra& (1,1) \quad \mbox{as} \; \lb\ra \infty,
\end{split}
\end{equation}
where $m_{\pm}=\lim_{\d\ra 0}m(\lb\pm i\d;x,t,\e)$.
The RH problem of finding the matrix $m(\lb)$ given $\nu(\lb)$
has a unique solution in the space $(1,1)+L^2(d\lb^{\frac{1}{2}})$.
The solution of the Cauchy problem (\ref{KdV})  is given by
\begin{equation}
\label{u(x)}
u(x,t,\e)=-2i\e\partial_x m_{11}(x,t,\e)
\end{equation}
where
\begin{equation}
m_1(\lb; x,t,\e)=1+m_{11}\lb^{-\frac{1}{2}}+O(\lb^{-1}), \quad \lb\ra
\infty,
\end{equation}
see again (\cite{S}, \cite{BDT}).
In this paper we study the Cauchy problem (\ref{KdV}) in the limit
$\e\ra 0$, the so called zero-dispersion limit of the KdV equation.
We use the WKB approximation with one turning point to  calculate the
reflection coefficient \cite{BO}
\begin{equation}
\nonumber
\begin{split}
r(\lb;\e)& \simeq -i e^{-2i\rho(\lb)/ \e}\chi_{[0,1]}(\lb)\\
\rho(\lb)&\simeq \lb^{\frac{1}{2}}x(\lb)-\int_{-\infty}^{x(\lb)}
[\lb^{\frac{1}{2}}-(\lb-u_0(x))^{\frac{1}{2}}]dx,\\
\end{split}
\end{equation}
where the quantity $x(\lb)$ is defined by the relation
$u_0(x(\lb))=\lb$. As usual $\chi_{[0,1]}(\lb)$ denotes the
characteristic
function of the interval $[0,1]$.
From the above considerations the jump matrix reduces to the identity
matrix for $\lb>1$ and our RH problem is reduced to the interval
$(-\infty, 1]$.
The quantity $\rho(\lb)$ can be expressed also in the form
\begin{equation}
\label{rho}
\rho(\lb)=\dfrac{1}{2}\int_{0}^{\lb}\dfrac{f(y)}{\sqrt{\lb-y}}dy
\end{equation}
where $f(u)|_{t=0}$ is the inverse function of the initial data
$u_0(x)$.
In the following we identify $r$ with its WKB approximation.

\noindent
Following the procedure in \cite{DVZ}, we introduce   a change of
the dependent variable $m$
\begin{equation}
\label{g}
M(\lb)=m(\lb)e^{\g(\lb)\sg_3/\e}
\end{equation}
where the scalar function $\g(\lb)=\g(\lb;x,t)$
is analytic in $\lb$ off the line $(-\infty, 1]$ and satisfies
$\g_+(\lb)+\g_-(\lb)=0$, for $\lb\in(-\infty,0)$ and $\g(\lb)\ra\ 0$
as $\lb\ra \infty$.
From (\ref{g}) and (\ref{u(x)}) we obtain
\begin{equation}
\nonumber
u(x,t)=-2i\e \partial_x M_{11}(x,t)-2\partial_x \g_1(x,t),
\end{equation}
where $\g(\lb)= \g_1(x,t)/\sqrt\lb+O(1/\lb)$ and $M_1(\lb;x,t)=1+
M_{11}(x,t)/\lb^{\frac{1}{2}}$.
The RH problem in the new variable becomes
$M_+(\lb)=M_-(\lb)\Up(\lb)$, $\lb\in(0,1)$,
where
\begin{equation}
\nonumber
\Up=\begin{pmatrix}
0& -ie^{-ih/\e}\\
-ie^{ih/\e}&e^{-i(\g_+-\g_-)/\e}\\
\end{pmatrix}
\end{equation}
and  $h=\g_++\g_--2\rho+2\alpha$.
For computing the small $\e$ asymptotics we follow the technique
in \cite{DVZ}.
The interval  $0<\lb<1$  is partitioned into finitely many intervals 
$I_j=(u_{2j},u_{2j-1})$, $\,j=1,\dots,g+1$ and 
$0=u_{2g+2}<u_{2g+1}<\dots<u_2<u_1<1$, $g\geq 0$. Using the steepest
descent method \cite{DVZ},\cite{DZ},
  the jump matrix  $\Up$ can be reduced
to one of the two forms with exponentially small errors  as $\e\searrow
0$,
\beqa
\begin{array}{lll}
&a0)&\quad \begin{pmatrix}
0&-ie^{-ih/\e}\\
-ie^{ih/\e}&0\\
\end{pmatrix}\quad \lb\in\cup_{j=1}^{g+1}I_j,\\
&&\\
&b0)&\quad \begin{pmatrix}
1&0\\
0&1\\
\end{pmatrix},\quad \lb\in(0,1)\backslash\cup_{j=1}^{g+1}I_j.\\
\end{array}
\eeqa
The function $\g$ satisfies the following conditions:
\begin{equation}
\label{RHg1}
\dfrac{(\g_+-\g_-)}{2i}<0,\quad \mbox{ and}\; h^{\prime}(\lb)=0,\;\;\mbox{thus}\;\;
\g_++\g_--2\rho+2\a=-\Om_j,\quad \lb\in\cup_{j=1}^{g+1}I_j,
\end{equation}
 where $\Om_j$ is some constant of integration;
\vskip 0.3cm
\begin{equation}
\label{RHg2}
 \g_+-\g_-=0 \quad \mbox{ and}\;\; h^{\prime}(\lb)>0,\quad \lb\in(0,1)\backslash\cup_{j=1}^{g+1}(u_{2j},u_{2j-1}).
\end{equation}
The remaining RH problem is the following
\beqa
\label{RH1}
\begin{array}{lcl}
M_+&=&M_-\sg_1, \quad \lb\in(-\infty, 0),\\
&&\\
M_+&=&-i M_-\sg_1 e^{-i\sg_3\Om_j/\e}, \quad
 \lb\in\cup_{j=1}^{g+1}I_j,\\
\end{array}
\eeqa
where $\sg_3=\begin{pmatrix} 1&0\\0&-1\\ \end{pmatrix}$.

To complete the solution of the initial value problem
(\ref{KdV})  we have to determine the phase function $\g$, the
 intervals $I_j$, $j=1,\dots,g+1$, the values $\Om_j$.  
Once all these steps have been made, we have
to solve the RH problem (\ref{RH1}), so that 
 the solution of the initial value
problem (\ref{KdV}) can be expressed by \cite{DVZ}, \cite{V1}
\[
u(x,t,\e)=\sum_{j=1}^{2g+1}u_j+2a_1^0-2\e^2\dfrac{\partial^2}{\partial
x^2}
\log\theta(\vec{\Omega}/(2\pi \e)),
\]
where $a_1^0$ has been defined in (\ref{b6}) and
$\vec{\Omega}=(\Omega_1,\Omega_2,\dots,\Omega_g)$. The theta
function is defined by 
\[
\theta(z,B)=\sum_{m\in Z^g}e^{2\pi i(m,z)+\pi i(m,Bm)}, \quad
z\in C^g,
\]
where $B$ is the period matrix of the holomorphic differentials (\ref{holo}),
namely
$B_{ij}=\int_{\beta_j}\phi_i$, $i,j=1,\dots,g$.

%%%%%%%%%%%%%%%%%%%%%%%%%%%%%%%%%%%%%%%%%%%%%%%%%%%%%%%%%%%%%%%%%%%%%%%%%%%%%%%%%%%%%%%%%%%%%%%%%%%%%%%%%%%%%%%%%%%%%%%%%%%%%%%%%%%%%%%%%%%%%%%%%%%%%%%%%%%%%%%%%%%%%%%%%%%%%%%%%%%%%%%%%%%%%%%%%%%%%%%%%%%%%%%%%%%%%%%%%%%%%%%%%%%%%%%%%%%%%%%%%%%%%%%%%%%%%%%%%%%%%%%%%%%%%%%%%%%%%%%%%%%%%%%%%%%%%%%%%%%%%%%%%%%%%%%%%%%%%%%%%%%%%%%%%%%%%%%%%%%%%%%%%%%%%%%%%%%%%%%%%%%%%%%%%%%%%%%%%%%%%%%%%%%%%%%%
\section{Determination of $\g(\lb)$}
We observe that for each fix $x$ and $t$ the function
$\g^{\prime}(\lb)$ satisfies the following RH problem
\begin{equation}
\label{RHg0}
\begin{aligned}
\g^{\prime}_++\g^{\prime}_-=&0, \quad \lb\in(-\infty,0)\\
%\label{RHg2}
\g^{\prime}_+-\g^{\prime}_-=&0, \quad \lb>1\; \mbox{and}\; \lb\in
(0,1)-\li_g,\\
%\label{RHg3}
\g^{\prime}_++\g^{\prime}_--2\rho^{\prime}+2\alpha^{\prime}=&0, \quad
\lb\in\li_g,\\
\end{aligned}
\end{equation}
where $\li_g=\cup_{j=1}^{g+1}I_j$. We call the intervals $I_j$  bands,
while the intervals  $(u_{2j+1},u_{2j})$, $j=0,\dots,g$, $u_0=1$,
 are  called gaps. 
We add the requirement that
\begin{equation}
\label{c1}
(\sqrt{\lb}\g^{\prime}(\lb))_{\pm},\;\;
\mbox{are continuous  functions for real} \;\lb.
\end{equation}
 It follows that 
$\g^{\prime}(\lb)_{\pm}$ are continuous functions for $\lb\in(0,1)$.

We recall that the condition $\g(\lb)=O(\lb^{-\frac{1}{2}})$ for large
$\lb$, implies
\begin{equation}
\label{c2}
\g^{\prime}(\lb)=O(\lb^{-\frac{3}{2}}).
\end{equation}
We observe that the condition (\ref{RHg2}) implies the following normalization
condition
\begin{equation}
\label{normalization}
\int_{ \alpha_j}\g^{\prime}(\lb)d\lb=0,\quad j=1,\dots,g,
\end{equation}
where $\alpha_j$ is any clockwise close  loop around the cut $I_j$.
When the loops $\alpha_j$ collapse to the interval $I_j$ described
twice,  the conditions (\ref{normalization}) become
\begin{equation}
\label{c3}
\int_{I_j}(\g^{\prime}_+-\g^{\prime}_-)d\lb=0, \quad j=1,\dots,g.
\end{equation}

The solution of the RH problem (\ref{RHg0})  that satisfies (\ref{c1})
is given by the integral
\cite{Mu}
\begin{equation}
\label{g1}
\g^{\prime}(\lb)=\dfrac{y(\lb)}{2 \pi i}\int_{\li_g}\dfrac{(2\rho^{\prime}(z)-2\alpha^{\prime}(z))dz}{(z-\lb)\yz}
\end{equation}
where
\begin{equation}
\label{yl}
y^2=\prod_{j=1}^{2g+1}(\lb-u_j)\,.
\end{equation}
We choose $y(\lb)$ to be analytic off the intervals  $(-\infty,0]\cup\li_g $,
   real and positive for $\lb>u_{1}$ and we
 denote
$y^+(\lb)$ the boundary value from above the cut $(-\infty,0]\cup\li_g $. 

In order for (\ref{g1}) to satisfy (\ref{c2})  we must impose
the following moment conditions
\begin{equation}
\label{moment}
\int_{\li_g}\dfrac{ \rho^{\prime}(\lb)-\a^{\prime}(\lb)) }{y^+(\lb)  }\lb^kd\lb=0,\quad
k=0,\dots,g.
\end{equation}
Furthermore we must impose the 
normalization conditions (\ref{normalization}).
We observe that (\ref{normalization}) and (\ref{moment}) represent a
system of $2g+1$ algebraic equations which, in principle, determines the
end-points
$u_1,u_1,\dots,u_{2g+1}$ of the intervals $I_j$, $j=1,\dots,g+1$.

\noindent
For $g=0$ the moment  condition reduces to the form
\[
\dfrac{2}{\pi i}\int_{0}^u\dfrac{ \rho^{\prime}(\lb)-\a^{\prime}(\lb))
}{\sqrt{\lb-u  }}d\lb=f(u)-6tu-x=0,
\]
which is the solution of the zero-phase equation (\ref{zeroph}).

\noindent
We observe that
\[
\g^{\prime}_+(z_0)-\g^{\prime}_-(z_0)=\g(z_0)=\dfrac{y(z_0)}{\pi
i}\int_{\li_g}\dfrac{(2\rho^{\prime}(z)-2\alpha^{\prime}(z))dz}{(z-z_0)\yz}\,
\quad z_0\in\li_g 
\]
and
\[
\g^{\prime}_+(z_0)-\g^{\prime}_-(z_0)=0,\quad \lb\in(0,1)\backslash\cup_{j=1}^{g+1}(u_{2j},u_{2j-1}).
\]
Therefore because of the assumption of the
 continuity of $\g^{\prime}_{\pm}$ on $(0,1)$, the following relation must be satisfied
\begin{equation}
\label{c11}
\lim_{z_0\ra u_i}\g^{\prime}(z_0)=0,\quad i=1,\dots,2g+1,
\end{equation}
where $z_0\in(u_{i+1},u_i)$ for $i$ odd and $z_0\in(u_{i-1},u_i)$ for
$i$ even. 

\noindent
We consider  analytic initial data or smooth  initial data in $(0,1)$
that
satisfies (\ref{const}), 
 so that the function $\rho^{\prime}(\lb)$  is  H\"older continuous in
subsets of $(0,1)$,  namely 
\[
|\rho^{\prime}(\lb_1)-\rho^{\prime}(\lb_2)|<c|\lb_1-\lb_2|^{\d},\quad
 \forall
\lb_1,\lb_2\in J
\]
where the constant $c>0$, $0<\d\leq 1$, and $J$ is some open subset of $(0,1)$.
\begin{lem}\cite{Mu}
\label{Mu}
If the function $\rho^{\prime}(\lb)$ is H\"older continuous near and at
 $u_i$, $i=1,\dots,2g+1$, then 
\[
\lim_{z_0\ra u_i}\g^{\prime}(z_0)=0,\quad i=1,\dots,2g+1,
\]
where $z_0\in(u_{i+1},u_i)$ for $i$ odd and $z_0\in(u_{i-1},u_i)$ for
$i$ even.
\end{lem}
The above lemma guarantees the consistency of the  
  assumptions (\ref{c1}) and (\ref{c11}).

\noindent
These considerations suggests to
 build a second solution $\tilde{\g}^{\prime}(\lb)$ of the RH
problem  (\ref{RHg0}) requiring that $\gtp$ satisfies (\ref{c2}) and
(\ref{c3}) while we impose the continuity (\ref{c1}) as constraint.
The solution $\gtp$ is given by the expression
\begin{equation}
\label{gtp}
\gtp=\dfrac{1}{(2\pi i)\yl}\int_{\li_g}\dfrac{2\yz (\rho^{\prime}(z)-2\a^{\prime}(z))dz}{z-\lb}-\dfrac{Q_{g-1}(\lb)}{\yl}.
\end{equation}
The polynomial $Q_{g-1}(\lb)$ has degree $g-1$ and its coefficients are
uniquely determined from (\ref{c3}) or (\ref{normalization}).
It is easy to verify that
\begin{equation}
\dfrac{Q_{g-1}(\lb)}{\yl}=\dfrac{1}{2\pi i}
\sum_{k=1}^g\dfrac{\phi_k(\lb)}{d\lb}
\int_{\li_g}dz (2\rho^{\prime}(z)-2\a^{\prime}(z))
\int_{\a_k}\dfrac{\yz d\eta}{y(\eta)(z-\eta)},
\end{equation}
where  $\phi_k(\lb)$, $k=1,\dots g$  is the basis of holomorphic
differential defined in (\ref{holo}).

\begin{lem}
The function $\gtp$ satisfies  the conditions (\ref{c2}) and (\ref{c3}).
\end{lem}
The continuity on the function $\gtp$ is obtained imposing that
the end points $u_1,\dots, u_{2g+1}$ evolve according to the
equations
\begin{equation}
\label{zerog}
\gtp_{\lb=u_i}=0, \quad  i=1,\dots,2g+1.
\end{equation}
The next theorem establishes the equivalence between the two different
solutions $\g^{\prime}(\lb)$ and $\gtp$  of the RH (\ref{RHg0}).
\begin{theo}
\label{equiv}
The function function $\gtp$ and the set of algebraic equations
 (\ref{zerog}) is equivalent to the function
 $\g^{\prime}(\lb)$ and the set of algebraic  equations (\ref{moment}) and
(\ref{normalization}). The equivalence is established for any
 $C^{\infty}$ initial data satisfying (\ref{const}).
\end{theo}
\proof
We write $\gtp$  in  the form
\begin{equation}
\label{o1}
\gtp=\dfrac{1}{\yl}\int_{\li_g}\dfrac{y^2(z)
(2\rho^{\prime}(z)-2\a^{\prime}(z))dz}{(2\pi i)\yz(z-\lb)}-
\sum_{l=1}^g\dfrac{\phi_l(\lb)}{ d\lb}\int_{\li_g}dz
\dfrac{(2\rho^{\prime}(z)-2\a^{\prime}(z))}{(2\pi i)\yz}
\int_{\a_l}\dfrac{y^2(z) d\eta}{y(\eta)(z-\eta)}.
\end{equation}
Using  the identity
\begin{equation}
\dfrac{z^k}{z-\lb}=\sum_{j=0}^{k-1}z^{j}\lb^{k-1-j}+
\dfrac{\lb^k}{z-\lb},
\end{equation}
we obtain
\begin{equation}
\label{switch}
\dfrac{y^2(z)}{z-\lb}=\dfrac{y^2(\lb)}{z-\lb}+\sum_{k=0}^{2g}(-)^k
s_{2g-k} \sum_{j=0}^{k}z^{j}\lb^{k-j}
\end{equation}
where the $s_k$'s  are the symmetric function in the variables
$u_1, u_2,\dots u_{2g+1}$, namely   $s_0=1$,
$s_1=\sum_{k=1}^{2g+1}u_k$, $s_2=\sum_{k<j}u_ku_j$  and so on.
Using (\ref{switch}) we can rewrite (\ref{o1}) in the form
\begin{equation}
\label{o2}
\begin{split}
\gtp=&\dfrac{\yl}{2\pi i} \int_{\li_g}\dfrac{\rhoap dz}
{\yz(z-\lb)}-
\sum_{l=1}^g \dfrac{\phi_l(\lb)}{(2\pi i)d\lb}\int_{\li_g}\dfrac{\rhoap}{\yz}dz
\int_{\a_l}\dfrac{y(\eta)}{z-\eta}d\eta\\
&+\dfrac{1}{(2\pi i)\yl}\sum_{k=0}^{2g}(-)^k
s_{2g-k} \sum_{j=0}^{k}\lb^{k-j} \int_{\li_g}\dfrac{\rhoap
z^{j}}{\yz}dz\\
-&\dfrac{1}{2 \pi i}\sum_{l=1}^g\dfrac{\phi_l(\lb)}{d\lb}\sum_{k=0}^{2g}(-)^k
s_{2g-k} \sum_{j=0}^{k}\int_{\li_g}\dfrac{\rhoap z^{j}}{\yz}dz
\int_{\a_l}\dfrac{\eta^{k-j}}{y(\eta)}d\eta.
\end{split}
\end{equation}
Imposing (\ref{zerog}),
 we can see that the
first term in (\ref{o2})
 is automatically zero at the branch
points  by lemma~\ref{Mu}. Therefore, using (\ref{o2}), the equations (\ref{zerog}) imply
\begin{equation}
\nonumber
\begin{split}
\label{a1}
&\left[-\sum_{l=1}^n\sum_{m=0}^{n-1}\ga_l^m \lb^m\int_{\li_g}
\dfrac{\rhoap }{\yz}dz
\int_{\a_l}\dfrac{y(\eta)}{z-\eta}d\eta\right.\\
&+\sum_{k=0}^{2g}(-)^k
s_{2g-k} \sum_{j=0}^{k}\lb^{k-j} \int_{\li_g}\dfrac{\rhoap
z^{j}}{\yz}dz\\
&\left.\left.-\sum_{l=1}^g\sum_{m=0}^{g-1}\ga_l^m \lb^m\sum_{k=0}^{2g}(-)^k
s_{2g-k} \sum_{j=0}^{k}\int_{\li_g}\dfrac{\rhoap z^{j}}{\yz}dz
\int_{\a_l}\dfrac{\eta^{k-j}}{y(\eta)}d\eta\right]\right|_{\lb=u_i}=0,
\end{split}
\end{equation}
for $ i=1,\dots,2g+1.$
The above quantity is a polynomial in the $\lb$ variable
 of degree $2g$ that  must  have
$2g+1$ zeros, therefore it is identically zero.
From the coefficients of degree $g$ to $2g$ we get the moments
conditions (\ref{moment}).
For the coefficients of degree $m$, $m=0,\dots,g-1$ we get the relations
\begin{equation}
\nonumber
\begin{split}
\label{a2}
&-\sum_{l=1}^g\ga_l^m \int_{\li_g}\dfrac{\rhoap}{\yz}dz
\int_{\a_l}\dfrac{y(\eta)}{z-\eta}d\eta
+\sum_{k=m+g+1}^{2g}(-)^k
s_{2g-k}  \int_{\li_g}\dfrac{\rhoap
z^{k-m}}{\yz}dz\\
&-\sum_{l=1}^g\ga_l^m \sum_{k=g+1}^{2g}(-)^k
s_{2g-k} \sum_{j=g+1}^{k}\int_{\li_g}\dfrac{\rhoap z^{j}}{\yz}dz
\int_{\a_l}\dfrac{\eta^{k-j}}{y(\eta)}d\eta\equiv 0, \quad m=0,\dots,g-1.
\end{split}
\end{equation}
Using (\ref{snorm})  the above relation simplifies to the form
\begin{equation}
\label{a3}
\sum_{l=1}^g \ga_l^m\int_{\li_g}\dfrac{\rhoap}{\yz}dz
\int_{\a_l}\dfrac{y(\eta)}{z-\eta}d\eta \equiv 0, \quad m=0,\dots,g-1.
\end{equation}
Because the matrix $\{\ga_l^m\}$, $l=1,\dots,g$, $m=0,\dots,g-1$
is invertible, the relation (\ref{a3}) is equivalent to
\begin{equation}
\label{a4}
\int_{\li_g}\dfrac{\rhoap}{\yz}dz
\int_{I_l}\dfrac{y(\eta)}{z-\eta}d\eta \equiv 0, \quad l=1,\dots,g
\end{equation}
which coincides with the normalization conditions  (\ref{normalization}).
Therefore on the solution of (\ref{moment}) and (\ref{normalization})
the function $\gtp$ reads
\begin{equation}
\gtp=\dfrac{\yl}{2\pi i} \int_{\li_g}\dfrac{\rhoap dz}
{\yz(z-\lb)}=\g^{\prime}(\lb).
\end{equation}
\hfill $\square$

\noindent
In the same way we can prove that   $\g^{\prime}(\lb)$, the moment
conditions (\ref{moment}) and the normalization conditions
 (\ref{normalization}) are equivalent to  $\gtp$ and the equations
(\ref{zerog}).
We remark that the equivalence between the two solutions
 does not depend on the
fact that the problem has a priori a unique solution as follows
from the Lax-Levermore theory but it derives only on the structure
of the RH problem for the scalar function $\g$.

\noindent
We observe 
 that using $\dozl$,  the meromorphic analogue of the Cauchy kernel
 defined  in (\ref{CK}),
 we can write  $\gtp d\lb$ in the form
\begin{equation}
\gtp d\lb=-x\dpt-t\dqt+\Om^g(\lb),
\end{equation}
where 
\begin{equation}
\label{dpt}
\begin{split}
\dpt=&-\dfrac{1}{2\pi i}\int_{\li_g}\dozl z^{-\frac{1}{2}}\\
\dqt=&-\dfrac{12}{2\pi i}\int_{\li_g}\dozl z^{\frac{1}{2}}\\
\end{split}
\end{equation}
and
\begin{equation}
\label{OMp}
\Om^g(\lb)=-\pie\int_{\li_g}2\dozl \rho^{\prime}(z).
\end{equation}
Here and below the integrals in the $z$ variable are taken on the upper
side of  $\li_g$.

\noindent
Using the above representation we compute
the constants $\Omega_k$ defined in (\ref{RHg1}) 
\begin{equation}
\label{OMk}
\Omega_k=\int_{\beta_k}\gtp d\lb=-\int_{\beta_k}\int_{\li_g}\dozl\dfrac
{\rhoap}{2\pi i}dz
\quad k=1,\dots,g.
\end{equation}
\noindent
Integrating by parts the above identity and using (\ref{dozl}) and
(\ref{periods}) we obtain
\[
\Omega_k=4\int_{\li_g}\phi_k(z)((\rho(z)-\alpha(z)) dz,\quad k=1,\dots,g. 
\]
From the above it follows that $\Om_{g+1}=0$.

\noindent
The following theorem due to Krichever \cite{K} connects the solution
of the set of algebraic equations (\ref{moment}) and
(\ref{normalization}) or (\ref{zerog}), to a solution of the
Whitham equations.
\begin{theo}\cite{K}
Let us suppose that the $u_i$'s depend on $x$ and $t$ in such a way
that the conditions (\ref{zerog}) are fulfilled.
Then  $u_i=u_i(x,t)$, $i=1,\dots, 2g+1$, satisfies the
Whitham equations
\begin{equation}
\dfrac{\partial}{\partial
t}u_i=v_i(\vu)\dfrac{\partial}{\partial x}u_i, \quad i=1,\dots
2g+1,
\end{equation}
where
\begin{equation}
v_i(\vu)=\left.\dfrac{\dqt}{\dpt}\right|_{\lb=u_i},\quad
i=1,\dots,2g+1
\end{equation}
and the differentials  $\dpt$ and $\dqt$ have been defined in (\ref{dpt}).
\end{theo}

We remark that from proposition (\ref{propog}) the
 following identity is
easily verified
\begin{equation}
\label{vi}
v_i(\vu)=\left.\dfrac{\dqt}{\dpt}\right|_{\lb=u_i}=
\left.\dfrac{dq^g(\lb)}{dp^g(\lb)}\right|_{\lb=u_i}\quad
i=1,\dots,2g+1,
\end{equation} 
where $dp^g(\lb)$ and $dq^g(\lb)$ have been defined in (\ref{dp}).
The second expression of the $v_i(\vu)$'s in (\ref{vi}) is the classical
formula for the speeds of the Whitham equations obtained in \cite{FFM}. 

\noindent
We can write the algebraic equations (\ref{zerog}) in the form of the
so called hodograph transformation introduced by Tsarev \cite{T}:
\begin{equation}
\label{ch0}
x=-v_i(\vu)t+\tilde{w}_i(\vu), \quad i=1,\dots,2g+1,
\end{equation}
where
\begin{equation}
\label{twi}
\tilde{w}_i(\vu)=\left.\dfrac{\Om^g(\lb)}
{\dpt}\right|_{\lb=u_i},\quad i=1,\dots, 2g+1
\end{equation}
and  $\Om^g(\lb)$ has been defined in (\ref{OMp}).
As a consequence of theorem~\ref{equiv}, the set of algebraic equations
(\ref{moment}) and (\ref{normalization}) is equivalent to the
hodograph transformation (\ref{ch0}).

\noindent
In the next section we will show that the $\tilde{w}_i(\vu)$'s defined
in (\ref{twi}) coincide with the classical formulas provided in 
 \cite{FRT} or \cite{G1}. 

%%%%%%%%%%%%%%%%%%%%%%%%%%%%%%%%%%%%%%%%%%%%%%%%%%%%%%%%%%%%%%%%%%%%%%%%%%%%%%
%%%%%%%%%%%%%%%%%%%%%%%%%%%%%%%%%%%%%%%%%%%%%%%%%%%%%%%%%%%%%%%%%%%%%%%%%%%%%%%%%%%%%%%%%%%%%%%%%%%%%%%%%%%%%%%%%%%%%%%%%%%%%%%%%%%%%%%%%%%%%%%%%%%%%%%%%%%%%
%%%%%%%%%%%%%%%%%%%%%%%%%%%%%%%%%%%%%%%%%%%%%%%%%%%%%%%%%%%%%%%%%%%%%%%%%%%%%
%%%%%%%%%%%%%%%%%%%%%%%%%%%%%%%%%%%%%%%%%%%%%%%%%%%%%%%%%%%%%%%%%%%%%%%%%%%%%%%%%%%%%%%%%%%%%%%%%%%%%%%%%%%%%%%%%%%%%%%%%%%%%%%%%%%%%%%%%%%%%%%%%%%%%%%%%
\section{Solution of the Tsarev system and linear overdetermined
system of Euler-Poisson-Darboux type}
\setcounter{equation}{0}
We first define the Cauchy problem for the Whitham equations.
 The initial value problem consists of the
following. We consider the evolution on the $x-u$ plane
 of the initial curve $u(x,t=0)=u_0(x)$ according to the zero-phase
equation (\ref{zeroph}). 
The solution $u(x,t)$ of (\ref{zeroph}), with the initial 
data $u_0(x)$,  is given by the characteristic equation 
\begin{equation}
\label{zp}
x=-6tu+f(u)
\end{equation}
where $f(u)|_{t=0}$ is the inverse function of the initial data $u_0(x)$.
The solution $u(x,t)$ in (\ref{zp})  is globally well 
defined only for $0\leq t<t_0$, where 
$t_0=\frac{1}{6}\min_{u\in\Rn}[f^{\prime}(u)]$ is the time of gradient 
catastrophe of  (\ref{zp}). 
 Near the point of gradient catastrophe and for  a short time  $t>t_0$, 
 the evolving curve is given  by
 a multivalued function with
three branches $u_1(x,t)>u_2(x,t)>u_3(x,t)$, which
 evolve according to  the one-phase Whitham equations.

Outside  the multivalued region the solution is given by  the zero-phase
solution $u(x,t)$ defined in (\ref{zp}). On the
phase transition boundary  the zero-phase solution and the one-phase 
solution are attached $C^1$-smoothly.

\noindent
Since the Whitham equations are hyperbolic \cite{L}, other points 
of gradient catastrophe 
can appear in the  branches $u_1(x,t)>u_2(x,t)>u_3(x,t)$ themselves 
or in $u(x,t)$. 

In general, for $t>t_0$,  the evolving curve is    given  by
 a multivalued function with and odd number of branches
 $u_1(x,t)>u_2(x,t)>\dots>u_{2g+1}(x,t)$, $g\geq 0$. 
These branches evolve according to the $g$-phase Whitham equations. 
The $g$-phase solutions  for
{\it different $g$}  must be glued together in order to produce a $C^1$-smooth
curve in the $(x,u)$ plane evolving smoothly with $t$.  
The initial value problem of the Whitham equations is to determine, 
for almost all $t>0$ and $x$,  the phase
$g(x,t)\geq 0$ and the corresponding branches
 $u_1(x,t)>u_2(x,t)>\dots>u_{2g+1}(x,t)$
 from the initial data $x=f(u)|_{t=0}$. 

\noindent
The solution of the Whitham equations for a given $g$ is obtained by
the so called hodograph transformation introduced by Tsarev \cite{T}.
\begin{theo}
\label{tsarevt}
If $w_i(\vu)$, $\vu=(u_1,u_2,\dots,u_{2g+1})$, solves the linear
over-determined system
 \begin{equation}
\label{tsarev}
\dfrac{\partial w_i}{\partial u_j}=\dfrac{1}{v_i-v_j}\dfrac{\partial v_i}{\partial u_j}[w_i-w_j],\quad
i,j=1,2,\dots,2g+1,\;\;i\neq j,
\end{equation}
then the solution $(u_1(x,t),u_2(x,t),\dots,u_{2g+1}(x,t))$
of the hodograph transformation
\begin{equation}
\label{ch}
x=-v_i(\vu)\,t+w_i(\vu)
\,\quad i=1,\dots ,2g+1\,,
\end{equation}
satisfies system (\ref{whithamg}). Conversely, any solution
$(u_1,u_2,\dots, u_{2g+1})$  of (\ref{whithamg}) can be obtained in this
way in a neighborhood $(x_0,t_0)$ where the $u_{ix}$'s  are not vanishing..
\end{theo}

\noindent
To guarantee that the $g-$phase solutions for different $g$ are
attached continuously,    the following natural  boundary
conditions must be imposed on
  $w_i(u_1,u_2,\dots,u_{2g+1})$, $i=1,\dots,2g+1$, $g>0$.

\noindent
When $u_l=u_{l+1}$, $\;1\leq l\leq 2g$,
\begin{equation}
\label{b1}
w_l^g(u_1,\dots,u_l,u_l,\dots,u_{2g+1})=
w_{l+1}^g(u_1,\dots,u_l,u_l,\dots,u_{2g+1})
\end{equation}
and for $1\leq i\leq  2g+1,\;i\neq l,\,l+1$
\beqa
\label{b2}
%\begin{array}{lll}
w_i^g(u_1,\dots,u_l,u_l,\dots,u_{2g+1})=
w_i^{g-1}(u_1,\dots,\hat{u}_l,\hat{u}_l,\dots,u_{2g+1}).
\quad
%&w_i^g(u_1,\dots,u_l,u_l,\dots,u_{2g+1})=
%w_{i-1}^{g-1}(u_1,\dots,\hat{u}_l,\hat{u}_l,\dots,u_{2g+1}),
%\quad l+1<i\leq 2g+1\,
%\end{array}
\eeqa
The superscript $g$ and $g-1$  in the $w_i$'s
specify the corresponding genus
and the hat denotes the variable that have been dropped.
When $g=1$ and  $u_2=u_3$ we have that
\beqa
\label{b3}
\begin{array}{lll}
&w_1(u_1,u_3,u_3)=f(u_1)\\
  &w_2(u_1,u_3,u_3)=w_3(u_1,u_3,u_3)\,,
\end{array}
\eeqa
where $f(u)$ is the initial data. Similar conditions hold
true when $u_1=u_2$, namely
\beqa
\label{b4}
\begin{array}{lll}
&w_3(u_1,u_1,u_3)=f(u_3)\\
&w_1(u_1,u_1,u_3)=w_2(u_1,u_1,u_3).
\end{array}
\eeqa
We remark that the $v_i(\vu)$'s satisfy the boundary conditions
(\ref{b1}-\ref{b2}) and for $g=1$ we have
\[
v_1(u_1,u_3,u_3)=-6u_1,\quad  v_3(u_1,u_1,u_3)=-6u_3.
\]
The solution of the boundary value problem (\ref{tsarev}),
(\ref{b1}-\ref{b4})
has been obtained in \cite{G} for any smooth  monotonically increasing
initial data.
\begin{theo}\cite{G1}
\label{theoT1}
Let be $f(u)$ the inverse function of the smooth initial data $u_0(x,0)$.
If the function  $q_k=q_k(u_1,u_2,\dots,u_{2g+1})$, $1\leq k\leq g$,
 is the symmetric solution of the linear over-determined system
\beqa
\label{qk}
\left\{\begin{array}{lll}
2(u_i-u_j)\dfrac{\partial^2 q_k(\vu)}{\partial u_i\partial u_j}=
\dfrac{\partial q_k(\vu)}{\partial u_i}-\dfrac{\partial
q_k(\vu)}{\partial u_j},
\quad i\neq j,\;i,j=1,\dots, 2g+1,\;\;g>0 &&\\
&&\\
q_k(\underbrace{u,u,\dots,u}_{2g+1})=F_k(u)&&\\
&&\\
F_k(u)=\dfrac{2^{(g-1)}}{(2g-1)!!} u^{-k+\frac{1}{2}}
\dfrac{d^{g-k}}{du^{g-k}}
\left(u^{g-\frac{1}{2}}f^{(k-1)}(u)\right),&&
\end{array}\right.
\eeqa
with the ordering $1>u_1>u_2>\dots>u_{2g+1}>0$, then $w_i(\vu)$,
$i=1,\dots, 2g+1$, defined by
\begin{equation}
\label{wi}
w_i(\vu)=\dfrac{1}{P_0^g(u_i)}
\left[ 2\partial_{u_i}q_g(\vu)
\prod_{n=1,n\neq i}^{2g+1}(u_i-u_n)+
\sum_{k=1}^g q_k(\vu)
\sum_{n=1}^k(2n-1)\Gt_{k-n}P_{n-1}^g(u_i)\right],
\end{equation}
solves  the boundary value problem (\ref{tsarev}),
(\ref{b1}-\ref{b4}). Conversely every solution of (\ref{tsarev}),
(\ref{b1}-\ref{b4}) can be obtained in this way.
\end{theo}
In (\ref{wi}) the polynomials $P_n^g$'s have been defined in
(\ref{b6})
and the  $\Gt_{k}$'s are the coefficient of the expansion
for $\xi\rightarrow\infty$ of
\begin{equation}
\label{Gt}
y(\xi)=\xi^{g+\frac{1}{2}}(\Gt_0+\dfrac{\Gt_1}{\xi}+
\dfrac{\Gt_2}{\xi^2}+\dots+\dfrac{\Gt_l}{\xi^l}+\dots).
\end{equation}

The existence and uniqueness of the solution of the boundary value
problem (\ref{qk}) has been proved in \cite{FRT1}.
\begin{theo}\cite{FRT1}
 The solution of the boundary value problem (\ref{qk}) is unique,
  symmetric with respect to the variables $u_1,u_2,\dots,u_{2g+1}$ and reads
\begin{equation}
\label{fqk}
\begin{split}
q_k(\vu)=&\dfrac{1}{C}\int_{-1}^1\int_{-1}^1\dots\int_{-1}^1
 dz_1dz_2\dots dz_{2g}(1+z_{2g})^{g-1}(1+z_{2g-1})^{g-\frac{3}{2}}
\dots(1+z_{3})^{\frac{1}{2}}(1+z_{1})^{-\frac{1}{2}}  \times\\
&\\
&\dfrac{F_k(\frac{1+z_{2g}}{2}(\dots(\frac{1+z_{2}}{2}(
\frac{1+z_{1}}{2}u_1+\frac{1-z_{1}}{2}u_2)+\frac{1-
z_{2}}{2}u_3)+\dots)+\frac{1-z_{2g}}{2}u_{2g})}
{\sqrt{(1-z_1)(1-z_2)\dots(1-z_{2g})}},
\end{split}
\end{equation}
where $C=\prod_{j=1}^{2g}C_{j}$ and
\begin{equation}
\label{Crho}
C_{j}=\int_{-1}^1 \dfrac{(1+\mu)^{\frac{j}{2}-1}}{\sqrt{1-\mu} }d\mu\,.
\end{equation}
\end{theo}
The functions $F_k(u)$  and
the  solutions
$q_k(\vu)$, $k=1,\dots,g$,  of the boundary value
 problem (\ref{qk})
 satisfy the
following relations:
\beqa
\label{relations}
\begin{array}{cll}
\partial_u F_k(u)&=&\dfrac{2g+1}{2}F_{k+1}(u)+u \partial_u F_{k+1}(u),\quad
k=1,\dots, g-1,\quad g>0,\\
&&\\
\partial_{u_i}q_k(\vu)&=&\dfrac{1}{2}q_{k+1}(\vu)+u_i \partial_{u_i}
q_{k+1}(\vu) \quad i=1.\dots,2g+1,\quad k=1,\dots,g-1\,\quad g>0.
\end{array}
\eeqa

\begin{lem}\cite{G1}
\label{theoT2}
The solution of the Whitham equations described by
(\ref{ch}) where the $w_i(\vu)$'s  are given by (\ref{wi}) is
$C^1$-smooth on the phase transition boundaries.
\end{lem}

%%%%%%%%%%%%%%%%%%%%%%%%%%%%%%%%%%%%%%%%%%%%%%%%%%%%%%%%%%%%%%%%%%%%%%%%%%
%%%%%%%%%%%%%%%%%%%%%%%%%%%%%%%%%%%%%%%%%%%%%%%%%%%%%%%%%%%%%%%%%%%%%%%%%
%%%%%%%%%%%%%%%%%%%%%%%%%%%%%%%%%%%%%%%%%%%%%%%%%%%%%%%%%%%%%%%%%%%%%%%%%

The next theorem shows the equivalence between the hodograph transformation
defined in (\ref{ch0}) and the one define in (\ref{ch})

\begin{theo}
\label{wiequiv}
For any smooth monotonically increasing initial data satisfying
(\ref{const}),
the following identity is satisfied
\begin{equation}
\tilde{w}_i(\vu)\equiv w_i(\vu), \quad i=1,\dots 2g+1,
\end{equation}
where  the $\tilde{w}_i(\vu)$'s are defined in (\ref{twi}) and
 the $w_i(\vu)$'s are defined in (\ref{wi}).
\end{theo}

Therefore  combining theorem~\ref{equiv} and  and
theorem~\ref{wiequiv}, we deduce that  the hodograph transformation
(\ref{ch})
is equivalent to the set of algebraic equations (\ref{moment}) and
(\ref{normalization}).

The next theorem  shows  that  the hodograph transformation (\ref{ch})
 can be written
in a nice algebraic form. This is the first step to transform  the moment
 conditions
(\ref{moment}) and the normalization conditions 
(\ref{normalization})  into a combination of solutions of
 linear-overdetermined
systems of Euler-Poisson-Darboux type.
\begin{theo}
\label{hodoalge}
For $g>0$ the hodograph transformation (\ref{ch}) where the $w_i(\vu)$'s are defined in
(\ref{wi}) is equivalent to the following set of $2g+1$  algebraic equations
\begin{equation}
\begin{split}
\label{zero1}
&\sum_{j=1}^{2g+1}\partial_{u_j}q_{g-k}(\vu)-kq_{g-k+1}(\vu)=0,
\quad  k=0,\dots,g-2\\
&\sum_{j=1}^{2g+1}\partial_{u_j}q_{1}(\vu)-(g-1)q_{2}(\vu)-6\,t=0,\\
&2\sum_{j=1}^{2g+1}u_j\partial_{u_j}q_{1}(\vu)+q_{1}(\vu)-x-6\,t\sum_{j=1}^{2g+1}u_j=0,\\
\end{split}
\end{equation}
\begin{equation}
\label{zero2}
\int_{u_{2k}}^{u_{2k-1}}y(\lb)\Phi(\lb;\vu)d\lb=0,\quad k=1,\dots, g
\end{equation}
where the function $\Phi(\lb;\vu)$ is given by the relation
\begin{equation}
\label{Phi}
\Phi^g(\lb;\vu)=\partial_{\lb}\Psi^g(\lb;\vu)+\sum_{i=1}^{2g+1}
\partial_{u_i}\Psi^g(\lb;\vu).
\end{equation}
The function $\Psi^g(\lb;\vu)$
 satisfies the linear overdetermined system
of Euler-Poisson-Darboux type
\beqa
\label{bpsi}
\left\{\begin{array}{lll}
&&\dfrac{\partial}{\partial u_i}\Psi^g(\lb;\vu)-\dfrac{\partial}{\partial u_j}
\Psi^g(\lb;\vu)=2(u_i-u_j)\dfrac{\partial ^2}{\partial u_i\partial u_j}
\Psi^g(\lb;\vu),\quad i\neq j,\;\;i,j=1,\dots 2g+1\\
&&\dfrac{\partial}{\partial \lb}\Psi^g(\lb;\vu)-2\dfrac{\partial}{\partial u_j}
\Psi^g(\lb;\vu)=2(\lb-u_j)\dfrac{\partial ^2}{\partial \lb\partial u_j}
\Psi^g(\lb;\vu),\quad \;j=1,\dots 2g+1\\
&&\Psi^g(\lb;\underset{2g+1}{\underbrace{\lb,\dots,\lb}})=
\dfrac{2^{g}}{(2g+1)!!}f^{(g)}(\lb)
\end{array}\right.
\eeqa
where $f^{(g)}(\lb)$ is the $g$th derivative  of the
smooth monotonically increasing initial data $f(u)$.
\end{theo}
The above boundary value problem can be  integrate  in the form
\begin{equation}
\label{psi}
\begin{split}
\Psi^g(\lb;\vu)=&\dfrac{1}{K}\int_{-1}^1\int_{-1}^1\dots\int_{-1}^1
 dz_2dz_2\dots dz_{2g+2}(1+z_{2g+2})^g(1+z_{2g+1})^{g-\frac{1}{2}}
\dots(1+z_3)^{\frac{1}{2}}\times\\
&\\
&\dfrac{ f^{(g)}(\frac{1+z_{2g+2}}{2}(\dots(\frac{1+z_{3}}{2}(
\frac{1+z_{2}}{2}\lb+\frac{1-z_{2}}{2}u_1)+\frac{1-
z_{3}}{2}u_2)+\dots)+\frac{1-z_{2g+2}}{2}u_{2g+1}) }
{ \sqrt{(1-z_2)(1-z_3)\dots(1-z_{2g+2})} },\\
\end{split}
\end{equation}
where $K=\prod_{j=2}^{2g+2}C_j$ and the $C_j$'s have been defined in (\ref{Crho}). The solution obtained is symmetric with respect to the variables
$u_1,\dots, u_{2g+1}$. A similar formula can be obtained for $\Phi^g(\lb;\vu)$.

In the next section we identify the system (\ref{zero1}) 
with the  moment conditions (\ref{moment}) and system (\ref{zero2}) 
with the normalization
condition (\ref{normalization}).

%%%%%%%%%%%%%%%%%%%%%%%%%%%%%%%%%%%%%%%%%%%%%%%%%%%%%%%%%%%%%%%%%
%%%%%%%%%%%%%%%%%%%%%%%%%%%%%%%%%%%%%%%%%%%%%%%%%%%%%%%%%%%%%%%%
%%%%%%%%%%%%%%%%%%%%%%%%%%%%%%%%%%%%%%%%%%%%%%%%%%%%%%%%%%%%%%%%%
\vskip 0.5cm

%%%%%%%%%%%%%%%%%%%%%%%%%%%%%%%%%%%%%%%%%%%%%%%%%%%%%%%%%%%%%%%%%%%%%%%%%%%%%%%
%%%%%%%%%%%%%%%%%%%%%%%%%%%%%%%%%%%%%%%%%%%%%%%%%%%%%%%%%%%%%%%%%%%%%%%%%%%%%
%%%%%%%%%%%%%%%%%%%%%%%%%%%%%%%%%%%%%%%%%%%%%%%%%%%%%%%%%%%%%%%%%%%%%%%%%%%%%%
%%%%%%%%%%%%%%%%%%%%%%%%%%%%%%%%%%%%%%%%%%%%%%%%%%%%%%%%%%%%%%%%%%%%%%%%%%%%%

\noindent
{\bf Proof of Theorem~\ref{wiequiv}}.

\noindent
We show that  the $\tilde{w}_i(\vu)$'s defined in (\ref{twi})
satisfy the Tsarev system
(\ref{tsarev}) and the boundary conditions (\ref{b1}-\ref{b4}).
Therfore by theorem~\ref{theoT1}, we obtain $\,\tilde{w}_i(\vu)=w_i(\vu)$. We
follow the steps in \cite{FRT},\cite{K}.

The quantities $\tw_i(u_1,u_2,\dots,u_{2g+1})$, $\;i=1,\dots 2g+1$,
are well defined. Indeed let us write $d\tilde{p}^g(\lb)$ and $\Om^g(\lb)$
 in the form  $d\tilde{p}^g(\lb)=\dfrac{\tilde{P}^g(u_i,\vu)}{y(\lb)}d\lb$ and 
$\Om^g(\lb)=\dfrac{\chi^g(\lb,\vu)}{y(\lb)}d\lb$ 
where
\begin{equation}
\label{rr}
\chi^g(\lb,\vu)=-\dfrac{1}{\pi i}\left[\int_{\li_g}\dfrac{y(z)}{\lb-z}\rho^{\prime}(z)
-\sum_{j=1}^g \lb^{g-j} \int_{\li_g}N^g_j(z)\rho^{\prime}(z)\right]\,
\end{equation}
and the  $N^g_j$'s have been defined in  (\ref{N}). Here and below all
the  integrals are taken on the upper side of $\li_g$.
Then
\begin{equation}
\begin{split}
\label{twi1}
\tw_i(u_1,u_2,\dots,u_{2g+1})&=\dfrac{\chi^g(u_i,\vu)}{\tilde{P}^g(u_i,\vu)}\\
&=-\dfrac{1}{\pi i\tilde{P}^g(u_i,\vu)}\left[\int_{\li_g} \dfrac{y(z)}{u_i-z}
\rho^{\prime}(z)-\sum_{j=1}^g  u_i^{g-j} \int_{\li_g}N_j(z)\rho^{\prime}(z)\right]\,.
\end{split}
\end{equation}
Next we show that the $\tw_i$'s satisfy the Tsarev system (\ref{tsarev}).
The differentials $\partial_{u_j}\Omega^g(\lb)$ and
$\partial_{u_j}\dpt$ 
 are normalized Abelian differentials of the second kind  with a  
pole at $\lb=u_j$ of second order. Because of (\ref{twi}) the differential
$\dfrac{\partial }{\partial u_j}\Om^g-\tw_j\dfrac{\partial
}{\partial u_j}\dpt$ is holomorphic and 
\begin{equation}
\nonumber
\begin{split}
0=&\partial_{u_j}\int_{\a_k}(\Om^g(\lb)-\tw_j\,\dpt),\\
0=&\int_{\a_k} \dfrac{\partial }{\partial u_j}\Om^g-\tw_j
\dfrac{\partial }{\partial u_j}\dpt,\quad k=1,\dots,g,
\end{split}
\end{equation}
that follows from (\ref{norm}).
Therefore
\begin{equation}
\label{Omdp}
(\partial_{u_j}\Om^g(\lb)-\tw_j\partial_{u_j}\dpt)\equiv 0
\end{equation}
because it is a holomorphic differential having all the  
$\alpha$-periods equal to zero.

From (\ref{Omdp}) we obtain
\begin{equation}
\label{pr}
\dfrac{\partial }{\partial u_j}\chi^g(\lb)-\tw_j
\dfrac{\partial }{\partial u_j}\tilde{P}^g(\lb)=-\dfrac{1}{2}\dfrac{\chi^g(\lb)-\tw_j \tilde{P}^g(\lb)}{\lb-u_j}\,,
\quad i=1,2,\dots,2g+1.
\end{equation}
We use the above identity to evaluate $\partial_j \tw_i(\vu)$.
From (\ref{twi1}) and (\ref{pr}) we obtain
\begin{equation}
\begin{split}
\dfrac{\partial }{\partial u_j} \tw_i(\vu)=&\dfrac{\partial}{\partial u_j}\dfrac{\chi^g(u_i,\vu)}{\tP(u_i,\vu)}\,,\quad i\neq j\\
=&\dfrac{\partial_{u_j}\chi^g(u_i,\vu)-\tw_j\partial_{u_j}\tP(u_i,\vu)}{\tP(u_i,\vu)}+(\tw_j-\tw_i)\dfrac{\partial_{u_j}\tP(u_i,\vu)}{\tP(u_i,\vu)}\\
=&(\tw_j-\tw_i)\dfrac{\partial_{u_j}\tP(u_i,\vu)}{\tP(u_i,\vu)}-\dfrac{1}{2}\dfrac{\chi^g(u_i,\vu)-\tw_j\tP(u_i,\vu)}{(u_i-u_j)\tP(u_i,\vu)}\\
=&(\tw_j-\tw_i)\dfrac{\partial_{u_j}\tP(u_i,\vu)}{\tP(u_i,\vu)}-
\dfrac{1}{2}\dfrac{\tw_i-\tw_j}{u_i-u_j},
\end{split}
\end{equation}
which shows that
\begin{equation}
\label{ttt}
\dfrac{1}{\tw_i-\tw_j}\dfrac{\partial \tw_i}{\partial
u_j}=-\dfrac{\partial_{u_j}\tP(u_i,\vu)}{\tP(u_i,\vu)}-\dfrac{1}{2}\dfrac{1}{u_i-u_j}.
\end{equation}
In particular the  above argument  also applied to $d\tilde{q}^g(\lb)$,
 $d\tilde{p}^g(\lb)$
and $v_i(\vu)$  therefore
we also have
\[
\dfrac{1}{v_i-v_j}\dfrac{\partial v_i}{\partial u_j}=-\dfrac{\partial_{u_j}\tilde{P}^g(u_i,\vu)}{\tilde{P}^g(u_i,\vu)}-\dfrac{1}{2}\dfrac{1}{u_i-u_j},
\]
which when combined with (\ref{ttt}) proves the Tsarev relation for
the
$\tw_i$'s.

\noindent
Next we show that the functions $\tilde{w}_i(u_1,u_2,\dots,u_{2g+1})$, $i=1,\dots,2g+1$,
satisfy the boundary conditions (\ref{b1}-\ref{b4}).

In the following we use the superscript $g$ to
  denote the corresponding genus of the quantities we are referring
  to.
We need to consider the behavior of the Abelian differential
 $\dozl$  when two
branch  points become coincident.
For the purpose let   be $u_l=v+\sqrt\d$, $u_{l+1}=v-\sqrt\d$ where
$0<\d\ll 1$.
The differential
  $\dozl=\om^g_z(\lb;u_1,\dots,u_{l-1},v-\sqrt\d,v+\sqrt\d,\dots,
  u_{2g+1})$ has the following expansion for $\d\ra 0$ when $l$ 
is odd   \cite{Fay}
\begin{equation}
\label{dolead}
\om^{g}_z(\lb,\d)=\om^{g-1}_z(\lb)+\dfrac{\delta}{2}\om^{g-1}_z(v)
\partial_v\omega_v^g(\lb)+O(\d^2),
\end{equation}
where  $\omega_v^{g-1}(\lb)$ is the normalized Abelian differential
of the third kind having first order poles at the points
$Q^{\pm}(z)=(z,\pm \tilde{y}(z))$ with residue $\pm 1$ respectively  and it is
defined on the Riemann surface
\begin{equation}
\label{Sg}
\cs_{g-1}: \tilde{y}^2=\prod_{\substack{k=1\\k\neq l, l+1}}^{2g+1}
(\lb-u_k).
\end{equation}
The differential $O(\d^2)/\d^2$ has poles at $\lb=v$
of order at most  $4$ and zero residue.
In formula (\ref{dolead}) the quantity
$\om^{g-1}_z(v)=\dfrac{\om^{g-1}_z(\lb)}{d\lb}|_{\lb=v}$.

\noindent
When $l$ is even  $\dozl$ has the following expansion \cite{Fay} for
$\d\ra 0$
\begin{equation}
\label{dotrail}
\om^{g}_z(\lb)\simeq  \om_z^{g-1}(\lb)-\dfrac{1}{\log \d}\,\om^{g-1}_v(\lb)
\int_{Q^-(v)}^{Q^+(v)}\om^{g-1}_z(z).
\end{equation}
The above expansion contains also terms of order $\sqrt\d/\log/\d$.
Using (\ref{dolead}) and (\ref{dotrail}) we can get the expansion
of the differentials $\Om^g(\lb)$ and  $\dpt$ when $u_l=v+\sqrt\d$
and $u_{l+1}=v-\sqrt\d$,
$1\leq l\leq 2g$.  Let be  $C_v$  the contour from
$v-\sqrt\d$  to $v+\sqrt\d$ on the upper sheet of $\cs_g$.
 When $l$ is odd
$\li_g=\li_{g-1}\cup C_v$ where
\[
\li_{g-1}=[0,u_{2g+1}]\cup\dots\cup
 [u_{l+3},u_{l+2}]\cup[u_{l-1},u_{l-2}]\cup\dots\cup[u_2,u_1]
\]
 is the corresponding contour
 defined on the Riemann
surface $\cs_{g-1}$ of genus $g-1$.
From (\ref{dolead}) we obtain the expansion of $\Om^g(\lb)=\Om^g(\lb;u_1,\dots,u_{l-1},v-\sqrt\d,v+\sqrt\d,\dots, u_{2g+1})$ as $\d\ra 0$, namely
\begin{equation}
\label{Omlead}
\begin{split}
\Om^{g}(\lb)&=\Om^{g-1}(\lb)+\dfrac{\d}{2} \partial_v\omega_v^{g-1}(\lb)
\Om^{g-1}(v)
-\dfrac{1}{\pi i}\int_{C_v}\om^{g}_z(\lb)\rho^{\prime}(z)dz+O(\d^2),
\end{split}
\end{equation}
where
\[
\Om^{g-1}(v)=-\dfrac{1}{ \pi i}\int_{\li_{g-1}}\om^{g-1}_z(v)\rho^{\prime}(z)dz.
\]

When $l$ is even  $\li_g=\li_{g-1}\backslash C_v$
where
\[
\li_{g-1}=[0,u_{2g+1}]\cup\dots\cup
 [u_{l+4},u_{l+3}]\cup[u_{l+2},u_{l-1}]\cup[u_{l-2},u_{l-3}]\cup\dots\cup[u_2,u_1]
\]
and from (\ref{dotrail}) we obtain the expansion of $\Om^g(\lb)$
\begin{equation}
\label{Omtrail}
\begin{split}
\Om^{g}(\lb)&\simeq-\dfrac{1}{ \pi
i}\int_{\li_{g-1}}\om^{g}_z(\lb)\rho^{\prime}(z)dz
+\dfrac{1}{ \pi i}\int_{C_v}\om^{g}_z(\lb,\d)\rho^{\prime}(z)dz\\
&\simeq \Om^{g-1}(\lb)+\dfrac{1}{\log \d} \om^{g-1}_v(\lb)\int_{Q^-(v)}^{Q^+(v)}\Om^{g-1}(z)
+\dfrac{1}{ \pi i}\int_{C_v}\om^{g}_z(\lb)\rho^{\prime}(z)dz.
\end{split}
\end{equation}
The same expansions applies to $\dpt$, therefore combining
(\ref{Omlead}) and (\ref{Omtrail}) we obtain
\begin{equation}
\label{ZA}
\left.\dfrac{\Om^{g}(\lb)}{d\tilde{p}^{g}(\lb)}\right|_{\left[\substack{\lb=u_i\\u_l=u_{l+1}=v}\right]}=
\left.\dfrac{\Om^{g-1}(\lb)}{d\tilde{p}^{g-1}(\lb)}\right|_{\lb=u_i},\quad
i\neq l,l+1,\;i=1,\dots 2g+1.
\end{equation}
From the above  identity it is clear that the boundary conditions (\ref{b2})
 are satisfied.
In order to evaluate  (\ref{ZA}) at the points $\lb=v\pm
\sqrt\d$, we need to do some extra work. Let us defined the quantity
\begin{equation}
\label{chi}
\chi^{g}(\lb)=y(\lb)\dfrac{\Om^g(\lb)}{d\lb}.
\end{equation}
When $u_l=v+\sqrt\d$
and $u_{l+1}=v-\sqrt\d$, $l$ odd, using (\ref{Omlead}) we obtain
\begin{equation}
\label{CHI0}
\begin{split}
\chi^{g}(\lb)=&(\lb-v)\chi^{g-1}(\lb)-\dfrac{y(\lb)}{\pi
i}\int_{C_v}\dozl \rho^{\prime}(z)dz\\
+&\dfrac{\d}{2}\Om^{g}(v)(
\partial_v\tilde{y}(v)-(\lb-v)\sum_{k=1}^{g-1}\lb^{g-1-k}\partial_v
N_k^{g-1}(v))+O(\d^2),\\
\end{split}
\end{equation}
where now $O(\d^2)/\d^2$ is a polynomial in $\lb$.
Using (\ref{dolead}) we obtain the following expansion of the second
term in the above equation
\begin{equation}
\nonumber
\begin{split}
\dfrac{y(\lb)}{\pi i}\int_{C_v}\dozl \rho^{\prime}(z)dz=&\dfrac{1}{\pi
i}\int_{C_v}\dfrac{y^+(z)}{\lb-z}\rho^{\prime}(z)dz+\dfrac{1}{\pi
i}\int_{C_v}\dfrac{z-v}{\sqrt{(z-v)^2-\d}}\tilde{y}^+(z)\rho^{\prime}(z)dz\\
-&\dfrac{(\lb-v)}{\pi i}\sum_{k=1}^{g-1}\lb^{g-1-k}\int_{C_v} N_{k}^{g-1}(z)
\rho^{\prime}(z)dz+O(\d)
\end{split}
\end{equation}
Therefore
\begin{equation}
\label{CHI1}
\begin{split}
\left.\left(\dfrac{y(\lb)}{\pi i}\int_{C_v}\dozl
 \rho^{\prime}(z)dz\right)\right|_{\lb=v\pm \sqrt\d}&=
-\dfrac{1}{\pi i }
 \int_{C_v}\tilde{y}^+(z)\rho^{\prime}(z)
\dfrac{\sqrt{z-v\pm\sqrt\d}}{\sqrt{z-v\mp\sqrt\d}}dz\\
-&\dfrac{1}{\pi}\int_{C_v}\dfrac{z-v}{\sqrt{\d-(z-v)^2}}\tilde{y}^+(z)\rho^{\prime}(z)dz\\
&\pm\dfrac{\sqrt\d}
{\pi i}\sum_{k=1}^{g-1}v^{g-1-k}\int_{C_v} N_{k}^{g-1}(z)
\rho^{\prime}(z)dz+O(\d)\\
=&\pm\sqrt\d\,\tilde{y}^+(v)\rho^{\prime}(v)+\Rc_1(\d),
\end{split}
\end{equation}
where $\lim_{\d\ra 0}\dfrac{\Rc_1(\d)}{\sqrt\d}=0$.
Combining (\ref{CHI0}) and (\ref{CHI1}) we obtain
\begin{equation}
\label{chilead}
\chi^{g}(v\pm\sqrt\d)=\pm\sqrt\d(\chi^{g-1}(v)-
\tilde{y}^+(v)\rho^{\prime}(v))+\Rc_1(\d).
\end{equation}
In the same way we can get the expansion for
$\tilde{P}^g(v\pm\sqrt\d)$, namely
\[
\tilde{P}^g(v\pm\sqrt\d)=\pm\sqrt\d(
\tilde{P}^{g-1}(v)-\dfrac{\tilde{y}^+(v)}{2\sqrt v})+O(\d),
\]
which, when combined with (\ref{chilead}) gives
\begin{equation}
\label{as1}
\dfrac{\chi^{g}(v\pm \sqrt\d)}{\tilde{P}^g(v\pm\sqrt\d)}=
\dfrac{\chi^{g-1}(v)-\tilde{y}^+(v)\rho^{\prime}(v)
}{\tilde{P}^{g-1}(v)-
\tilde{y}^+(v)/(2\sqrt{v})}+\Rc_2(\d),
\end{equation}
where $\lim_{\d\ra 0}\Rc_2(\d)=0$.

\noindent
Using (\ref{Omtrail})   we obtain the following expansion of
$\chi^{g}(\lb)$ when $ u_l=v+\sqrt\d$
and $u_{l+1}=v-\sqrt\d$, $l$ even,
\begin{equation}
\label{chitrail}
\nonumber
\begin{split}
\chi^{g}(\lb)&\simeq (\lb-v)\chi^{g-1}(\lb)+\dfrac{y(\lb)}{  \pi
i}\int_{C_v}\dfrac{\om^{g}_z(\lb)}{d\lb}\rho^{\prime}(z)dz\\
+&\dfrac{1}{\log
\d}(\tilde{y}(v)-(\lb-v)\sum_{k=1}^{g-1}\lb^{g-k}N^{g-1}_k(v))
\int_{Q^-(v)}^{Q^+(v)}\Om^{g-1}(z).
\end{split}
\end{equation}
From the above and (\ref{CHI1}) we can evaluate
\begin{equation}
\nonumber
\chi^{g}(v\pm \sqrt\d)\simeq
\pm\sqrt\d(\chi^{g-1}(v)+\tilde{y}(v)\rho^{\prime}(v))
-\dfrac{1}{\log
\d}(\tilde{y}(v)\pm \sqrt\d\sum_{k=1}^{g-1}v^{g-1-k}N^{g-1}_k(v))
\int_{Q^-(v)}^{Q^+(v)}\Om^{g-1}(z)
\end{equation}
and
\[
\tilde{P}^g(v\pm\sqrt\d)\simeq \pm\sqrt\d (\tilde{P}^{g-1}(v)+
\dfrac{1}{2\sqrt{v}} \tilde{y}(v))+\dfrac{1}{\log
\d}(\tilde{y}(v)\pm \sqrt\d\sum_{k=1}^{g-1}v^{g-1-k}N^{g-1}_k(v))
\int_{Q^-(v)}^{Q^+(v)}d\tilde{p}^{g-1}(z).
\]
Combining the above two  expansions we obtain
\begin{equation}
\label{as2}
\dfrac{\chi^{g}(v\pm \sqrt\d)}{\tilde{P}^g(v\pm\sqrt\d)}\simeq
\dfrac{\int_{Q^-(v)}^{Q^+(v)}\Om^{g-1}(z)}{
\int_{Q^-(v)}^{Q^+(v)}d\tilde{p}^{g-1}(z)}.
\end{equation}
 The relations (\ref{as1}) and (\ref{as2}) show that
the boundary conditions (\ref{b1}) are satisfied.
From (\ref{ZA}) we obtain
\begin{equation}
w_1(u_1,v,v)=\dfrac{\chi^{0}(u_1)}{\tilde{P}^{0}(u_1)}
=\dfrac{\int_0^{u_1}\dfrac{\rho^{\prime}(z)}{\sqrt{z-u_1}}dz}
{\int_0^{u_1}\dfrac{1}{2\sqrt{z-u_1}\sqrt{z}}dz}=f(u_1),
\end{equation}
which shows that the boundary condition (\ref{b3}) is satisfied.
Analogous considerations can be done for proving (\ref{b4}).
Theorem~\ref{wiequiv} is then proved.\hfill $\square$

%%%%%%%%%%%%%%%%%%%%%%%%%%%%%%%%%%%%%%%%%%%%%%%%%%%%%%%%%%%%%%%%%%%%%%%%%%%%%%%%%%%%%%%%%%%%%%%%%%%%%%%%%%%%%%%%%%%%%%%%%%%%%%%%%%%%%%%%%%%%%%%%%%%%%%%%%%%%%%%%%%%%%%%%%%%%%%%%%%%%%%%%%%%%%%%%%%%%%%%%%%%%%%%%%%%%%%%%%%%%%%%%%%%%%%%%%%%%%%%%%%%%%%%%%%%%%%%%%%%%%%%%%%%%%%%%%%%%%%%%%%%%%%%%%%%%%%%%%%%%%%%
\vskip 0.3cm
\noindent
{\bf Proof of Theorem~\ref{hodoalge}}.
Following the steps in \cite{G1} we consider the polynomial
\begin{equation}
\label{zeta}
Z^g(\lb):=-xP^g_0(\lb)-12tP^g_1(\lb)+R^{g}(\lb),
\end{equation}
where    $R^0(\lb)=f(u)$ and  $R^{g}(\lb)$, $g>0$,  is given by the expression
\begin{equation}
\label{R2g}
R^{g}(\lb)=2\sum_{k=1}^{2g+1}\partial_{u_k}q_g(\vu)
\prod_{l=1,l\neq k}^{2g+1}(\lb-u_l)
+\sum_{k=1}^gq_k(\vu)
\sum_{l=1}^k(2l-1)\Gt_{k-l}P_{l-1}^g(\lb),
\end{equation}
with the polynomials $P^g_l(\lb)$, $l\geq 0$,  defined in (\ref{b6})
and the functions
$q_k(\vu)$, $k=1,\dots, g$,  defined in (\ref{qk}).
Then the hodograph transformation (\ref{ch}) is equivalent,
for $g>0$, to the equation 
\begin{equation}
\label{zero}
Z^g(\lb)\equiv 0,\quad g>0.
\end{equation}
The proof of the above proposition is obtained observing that
 the $w_i(\vu)$'s  defined in (\ref{wi}) are given by the ratio
$w_i(\vu)=\dfrac{R^{g}(u_i)}{P_0^g(u_i)}$,
$i=1,\dots,2g+1$,  where $R^{g}(\lb)$ is the polynomial defined in  (\ref{R2g}).
Hence  we can write  the hodograph transformation (\ref{ch}) in the
form
\begin{equation}
\label{SOL3}
[-xP^g_0(\lb)-12tP^g_1(\lb)+R^{g}(\lb)]_{\lb=u_i}=0, \quad i=1,\dots,2g+1. \;\;
\end{equation}
For $g>0$,  $\;Z^g(\lb)$ is a polynomial of degree $2g$   and
because of (\ref{SOL3})
it must have at least $2g+1$ real zeros. Therefore it is identically
zero.
Putting equal to zero the first $g+1$  coefficients of the
polynomial $Z^g(\lb)$ and using repeatedly the identity (\ref{relations}) we
obtain (\ref{zero1}).

\noindent
Putting to zero  the coefficients  of $Z^g(\lb)$ of degree  $0$
to degree $g-1$ is equivalent to the equations
\begin{equation}
\label{zero22}
\int_{\a_k}\dfrac{Z^g(\lb)}{y(\lb)}d\lb=0,\quad k=1,\dots, g
\end{equation}
Indeed when (\ref{zero1}) is satisfied,
  the differential
$\dfrac{Z^g(\lb)}{y(\lb)}d\lb$ is an holomorphic differential and by
(\ref{zero22}) it has  all its
alpha-periods equal to zero. Therefore it is identically zero. Hence
 the last $g$ equations  reads
\begin{equation}
\begin{split}
0=&\int_{\alpha_k}\dfrac{Z^g(\lb)}{y(\lb)}d\lb=\int_{\alpha_k}\dfrac{-xP^g_0(\lb)-12tP^g_1(\lb)+R^{g}(\lb)}{y(\lb)}d\lb\\
\label{ppp}
=&\int_{\alpha_k}2\dfrac{\sum_{i=1}^{2g+1}\partial_{u_i}q_g(\vu)
\prod_{j=1,j\neq i}^{2g+1}(\lb-u_j)}{y(\lb)}d\lb,\quad k=1,\dots,g.\\
\end{split}
\end{equation}
In the third equality of (\ref{ppp})  we have used the fact that
\[
\int_{\alpha_k}\dfrac{P^g_l(\lb)}{y(\lb)}d\lb=\int_{\alpha_k}\sg_l^g(\lb)=0,\quad l\geq 0,\;k=1,\dots,g,
\]
 because of the
normalization conditions (\ref{norm2}).
 The function $\Psi^g(\lb;\vu)$ defined in (\ref{psi})  satisfies the relations
\begin{equation}
\label{psir}
\dfrac{\Psi^g(\lb;\vu)}{\lb-u_i}-\dfrac{\Psi^g(u_i;\vu)}{\lb-u_i}=
2\partial_{u_i}\Psi^g(\lb;\vu),\;\;\;
2\partial_{u_i}q_g(\vu)=\Psi^g(u_i;\vu).
\end{equation}
Using (\ref{psir}) we can rewrite  the last term in (\ref{ppp}) in the form
\begin{equation}
\nonumber
\begin{split}
0=&\int_{\alpha_k}2\dfrac{\sum_{i=1}^{2g+1}\partial_{u_i}q_g(\vu)
\prod_{j=1,j\neq i}^{2g+1}(\lb-u_j)}{y(\lb)}d\lb,\quad k=1,\dots,g,\\
=&2\int_{u_{2k}}^{u_{2k-1}}y(\lb)\left(\sum_{i=1}^g\dfrac{\Psi^g(\lb;\vu)}{\lb-u_i}-2\partial_{u_i}\Psi^g(\lb;\vu)\right)d\lb\\
=&-4\int_{u_{2k}}^{u_{2k-1}} y(\lb)\left(\partial_{\lb} \Psi^g(\lb;\vu)+
\sum_{i=1}^g\partial_{u_i}\Psi^g(\lb;\vu)\right)d\lb,\quad k=1,\dots,g,
\end{split}
\end{equation}
where the last equality has been obtained integrating by parts.
 Using the definition of
$\Phi^g(\lb;\vu)$  in (\ref{Phi}) we rewrite the above relation in the form
\begin{equation}
\label{normphi}
0=-4\int_{u_{2k}}^{u_{2k-1}} y(\lb)\Phi^g(\lb;\vu)d\lb,\quad k=1,\dots,g,
\end{equation}
which is equivalent to  (\ref{zero2}).
\hfill$\square$

%%%%%%%%%%%%%%%%%%%%%%%%%%%%%%%%%%%%%%%%%%%%%%%%%%%%%%%%%%%%%%%%%%%%%%%%%%%%%%%%%%%%%%%%%%%%%%%%%%%%%%%%%%%%%%%%%%%%%%%%%%%%%%%%%%%%%%%%%%%%%%%%%%%%%%%%%%%%%%%%%%%%%%%%%%%%%%%%%%%%%%%%%%%%%%%%%%%%%%%%%%%%%%%%%%%%%%%%%%%%%%%%%%%%%%%%%%%%%%%%%%%%%%%%%%%%%%%%%%%%%%%%%%%%%%%%%%%%%%%%%%%%%%%%%%%%%%%%%%%%%%%%%%%%%%%%%%%%%%%%%%%%%%%%%%%%%%%%%%%%%%%%%%%%%%%%%%%%%%%%%%%%%%%%%%%%%%%%%%%%%%%%%%%%%%%%%
\subsection{The  function $\g^{\prime}$ and linear-overdetermined
systems
of Euler Poisson Darboux type.}
In this section we show that the set of algebraic equations described
by the moment conditions (\ref{moment}) and the  normalization conditions
(\ref{normalization}) can be written in terms of solutions of linear overdetermined
system
of Euler-Poisson-Darboux type introduced in \cite{FRT1}.
\begin{theo} 
\label{algequiv}
For any monotonically increasing analytic  initial data 
 satisfying (\ref{const}),
the set of algebraic equations defined by the moment  conditions (\ref{moment})
is equivalent, for $g>0$,  to (\ref{zero1}); the set  of algebraic equations defined by
the normalization conditions  (\ref{normalization}) is  equivalent,
for $g>0$,  to (\ref{zero2}), namely
\begin{equation} 
\begin{split}
\label{ealgequiv} 
&\sum_{j=1}^{2g+1}\partial_{u_j}q_{g-k}(\vu)-kq_{g-k+1}(\vu)=\dfrac{1}{\pi i
}\int_{\li_g}\dfrac{(\rho^{\prime}(z)-a^{\prime}(z))z^k}{y^+(z)}dz,
\quad  k=0,\dots,g-2\\ 
&\sum_{j=1}^{2g+1}\partial_{u_j}q_{1}(\vu)-(g-1)q_{2}(\vu)-6\,t=\dfrac{1}{\pi
 i}\int_{\li_g}\dfrac{(\rho^{\prime}(z)-a^{\prime}(z))z^{g-1}}{y^+(z)}dz\\
&2\sum_{j=1}^{2g+1}u_j \partial_{u_j}q_{1}(\vu)+q_{1}(\vu)-x-6\,t\sum_{j=1}^{2g+1}u_j=\dfrac{2}{\pi i
}\int_{\li_g}\dfrac{(\rho^{\prime}(z)-a^{\prime}(z))z^g}{y^+(z)}dz\\
&\int_{u_{2k}}^{u_{2k-1}}2y^+(\lb) \Phi(\lb;\vu)d\lb=\int_{I_k}(\g^{\prime}_+(\lb)-\g^{\prime}_-(\lb))d\lb,\quad k=1,\dots, g.
\end{split} 
\end{equation} 
\end{theo} 
\proof we first consider the moment conditions. Since
$a^{\prime}(z)=6tz^{\frac{1}{2}}+\dfrac{x}{2}z^{-\frac{1}{2}}$
we have 
\begin{equation}
\begin{split}
\int_{\li_g}\dfrac{a^{\prime}(z)z^k}{y(z)}dz&=0, \quad
k=0,\dots g-2\\
\dfrac{1}{\pi i
 }\int_{\li_g}\dfrac{a^{\prime}(z)z^{g-1}}{y^+(z)}dz&=6t,
\quad \dfrac{2}{\pi  i} \int_{\li_g}\dfrac{a^{\prime}(z)z^{g}}{y^+(z)}dz=
x+6t\sum_{j=1}^{2g+1}u_j
\end{split}
\end{equation}
For analytic initial data
we can deform the integrals
\begin{equation}
\label{II}
\int_{\li_g}\dfrac{\rho^{\prime}(z)z^k}{y(z)}dz=
\dfrac{1}{2}\sum_{j=0}^g\int_{\tilde{\a}_j}\dfrac{\rho^{\prime}(z)z^k}{y(z)}dz,\quad k=0,\dots,g
\end{equation}
where $\tilde{\a}_0$ is a  close loop around the interval
 $[0,u_{2g+1})$
 passing
through zero and the $\tilde{\a_j}$ are close loops around the intervals
$I_j$ and within the domain of analyticity of the initial data.
We can deform the contours $\cup\tilde{\a_j}$ to a single contour
$\cc=\cup_{j=0}^g \tilde{\a_j}\cup_{j=1}^g\ga_j$ where the contours
$\ga_j$ are the closed contours  plotted in the figure
below.
\vskip 0.2cm
\noindent
\vbox{\hsize 440pt
\hskip 1cm\centering{\psfig{figure=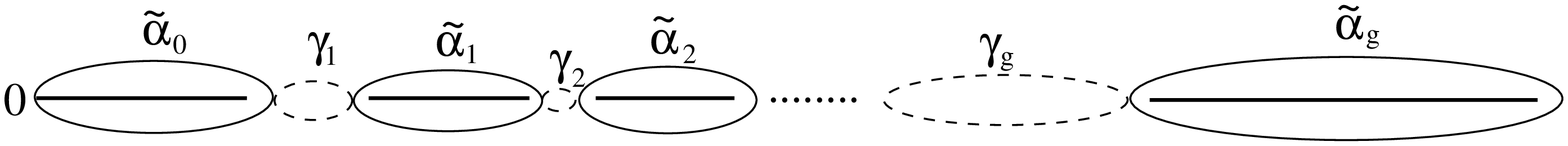,height=0.4in }}
}

\noindent
The integrals
\[
\int_{\ga_j}\dfrac{\rho^{\prime}(z)z^k}{y(z)}dz=0,\quad
j=1,\dots,g,
\]
therefore
\begin{equation}
\label{IIb}
\int_{\li_g}\dfrac{\rho^{\prime}(z)z^k}{y(z)}dz=\dfrac{1}{2}
\int_{\cc}\dfrac{\rho^{\prime}(z)z^k}{y(z)}dz,\quad k=0,\dots,g
\end{equation}
From the right hand side of (\ref{IIb}) it is 
 straightforward to verify that the
integrals $\int_{\li_g}\dfrac{\rho^{\prime}(z)z^k}{y(z)}dz$, $k\geq 0$, 
are symmetric with respect to the variables
$u_1,\dots, u_{2n+1}$ and
 satisfy the linear overdetermined system of
Euler-Poisson-Darboux type defined in (\ref{qk}) with the initial data
\begin{equation}
\label{III}
\begin{split}
&\left.\dfrac{1}{\pi i}\int_{\li_g}\dfrac{\rho^{\prime}(z)z^k}{y^+(z)}dz\right|_{[u_1=u_2=\dots=u_{2g+1}=u]}
=\partial_uF_{g-k}(u)-kF_{g-k+1}(u), \quad k=0,\dots, g-1,\\
&\left.\dfrac{2}{\pi i }\int_{\li_g}\dfrac{\rho^{\prime}(z)z^g}{y^+(z)}dz\right|_{[u_1=u_2=\dots=u_{2g+1}=u]}=2u\partial_uF_1(u)+F_1(u),
\end{split}
\end{equation}
where the functions $F_k(u)$, $k=1,\dots, g$,  have been defined in (\ref{qk}).
Because of the uniqueness of the solution of the boundary value problem
(\ref{qk}) the first $g+1$ equations in (\ref{ealgequiv}) are satisfied.

As regarding the normalization conditions (\ref{normalization})
 we have the following identity for
 $\lb\in I_j$, $j=0,\dots,g$ 
\begin{equation}
\label{Phi0}
\begin{split}
\g^{\prime}_+(\lb)-\g^{\prime}_-(\lb)=&
\dfrac{2y^+(\lb)}{\pi i}\left(v.p.\int_{\li_g}\dfrac{\rho^{\prime}(z)-a^{\prime}(z)}
{(z-\lb)y^+(z)}dz\right)\\
&=\dfrac{y^+(\lb)}{ \pi i}\sum_{j=0}^g\int_{\tilde{\a_j}}\dfrac{\rho^{\prime}(z)-a^{\prime}(z)}{(z-\lb)y^+(z)}dz\\
&=\dfrac{y^+(\lb)}{\pi i}\int_{\cc}\dfrac{\rho^{\prime}(z)-a^{\prime}(z)}{(z-\lb)y^+(z)}dz\\
&=2y^+(\lb)(\Phi^g(\lb;\vu)-6t\e_{g0}),
\end{split}
\end{equation}
where the function $\Phi^g(\lb;\vu)$ has been defined  in
(\ref{Phi}) and $\e_{g0}$ is equal to one for $g=0$ and zero
otherwise. 
The last identity in (\ref{Phi0}) has been obtained performing 
computations similar to the ones in (\ref{III}). 
Therefore
\begin{equation}
\label{n1}
0=\int_{I_j}(\g^{\prime}_+(\lb)-\g^{\prime}_-(\lb)d\lb=
2\int_{u_{2j}}^{u_{2j-1}} y^+(\lb)\Phi^g(\lb;\vu)d\lb,
\quad j=1,\dots,g, \;g>0
\end{equation}
and theorem~\ref{algequiv} in then proved. \hfill$\square$

\begin{theo}
The relations (\ref{ealgequiv}) are satisfied for smooth initial data.
\end{theo}
\proof the proof is obtained  combining  theorems \ref{equiv}, \ref{wiequiv}, \ref{hodoalge}
and \ref{algequiv}. \hfill$\square$
\vskip 0.1cm
\noindent
In the following we write the variational conditions in (\ref{RHg1})
and (\ref{RHg2}) in terms of the function $\Phi^g(\lb;\vu)$ defined in
(\ref{Phi}).
%%%%%%%%%%%%%%%%%%%%%%%%%%%%%%%%%%%%%%%%%%%%%
%%%%%%%%%%%%%%%%%%%%%%%%%%%%%%%%%%%%%%%%%%
\begin{theo}
The variational conditions (\ref{RHg1})
and
(\ref{RHg2}) can be written in the form
\begin{equation}
\label{RHg11}
0>\dfrac{\g_+(\lb)-\g_-(\lb)}{2i}=i\int_{\lb}^{u_{2j-1}}
y^+(\xi)(\Phi^g(\xi;\vu)-6t\e_{g0})d\xi,\quad \lb\in(u_{2j},u_{2j-1}),\;j=1,\dots,g+1
\end{equation}	
and
\begin{equation}
\label{lax0}
0<\g^{\prime}_++\g^{\prime}_--2\rho^{\prime}+2\a^{\prime}=2y(\lb)
(\Phi^g(\lb;\vu)-6t\e_{g0}),\quad
\lb\in(u_{2j+1},u_{2j}),\;j=0,\dots,g,\;u_0=1,
\end{equation}
where the function $\Phi^g(\lb;\vu)$  has been defined in (\ref{Phi})
and $\e_{g0}$ is equal to $1$ for $g=0$ and zero otherwise.
\end{theo} 
%%%%%%%%%%%%%%%%%%%%%%%%%%%%%%%%%%%%%%%%%%%%%%%%%%%%%%%%%%%%%%%%%%%%%%%%%
%%%%%%%%%%%%%%%%%%%%%%%%%%%%%%%%%%%%%%%%%%%%%%%%%%%%%%%%%%%%%%%%%%%%%%%%%%%%%
\proof  we first prove (\ref{RHg11}).  
From (\ref{Phi0}) and lemma~\ref{Mu} we obtain
\[
\dfrac{\g_+-\g_-}{2i}=-\int_{\lb}^{u_{2j-1}}
\dfrac{\g^{\prime}_+-\g^{\prime}_-}{2i}=i\int_{\lb}^{u_{2j-1}}
y(\xi)\Phi^g(\xi;\vu)d\xi<0,\quad 
\lb\in\cup_{j=1}^{g+1}(u_{2j},u_{2j-1}),
\]
 which coincides with (\ref{RHg11}).

As regarding (\ref{lax0}) we have
\begin{equation}
\label{GH}
\g^{\prime}_++\g^{\prime}_-=\dfrac{y(\lb)}{\pi
i}\int_{\li_g}
\dfrac{2\rho^{\prime}(z)}{y^+(z)(z-\lb)}dz-2\a^{\prime}(\lb)-12t\e_{g0}\,y(\lb),
\quad
\lb\in(u_{2j+1},u_{2j}),\;j=0,\dots,g,
\end{equation}
where $\e_{g0}$ is equal to one for $g=0$ and zero otherwise.

In order to express the integral 
\[
\int_{\li_g}
\dfrac{2\rho^{\prime}(z)}{y^+(z)(z-\lb)}dz
\]
as the  solution of  a linear overdetermined system of Euler-Poisson-Darboux
type, 
 let  us consider the function $q_{g+1}(\tilde{u_1},\tilde{u_2},u_1,u_2,
\dots,u_{2g+1})=q_{g+1}(\tilde{u_1},\tilde{u_2},\vu)$ which satisfies 
(\ref{qk}) with initial data
\[
q_{g+1}(\underbrace{u,u,\dots,u}_{2g+3})=\dfrac{2^{g}}{(2g+1)!!}f^{(g)}(u).
\]
Then 
\begin{equation}
\label{Philimit}
\left.\left(\sum_{j=1}^{2g+1}\partial_{u_j}q_{g+1}(\tilde{u_1},\tilde{u_2},\vu)+
\partial_{\tilde{u_1}}q_{g+1}(\tilde{u_1},\tilde{u_2},\vu)
+\partial_{\tilde{u_2}}q_{g+1}(\tilde{u_1},\tilde{u_2},\vu)\right)\right|_{\tilde{u_1}=\tilde{u_2}=\lb}=\Phi^g(\lb;\vu).
\end{equation}
The derivation of the above identity is straightforward.
Next we consider the first identity in (\ref{ealgequiv}) for the
function $q_{g+1}(\tilde{u_1},\tilde{u_2},\vu)$. Let us suppose
$ u_{2j}>\tilde{u_1}>\tilde{u_2}>u_{2j+1}$, for $j=0,\dots, g$, $u_0=1$,
let us define the interval $\tilde{I}=(\tilde{u_2},\tilde{u_1})$ 
and $Y(\lb)=y(\lb)\sqrt{(z-\tilde{u_2})(z-\tilde{u_1})}$. The function
$Y(\lb)$ is analytic in the complement of $(-\infty,0]\cup\li_g\cup I$ and 
 positive for $\lb>u_1$. 
$Y^+(\lb)$ denotes  the boundary value from above.  Then 
by (\ref{ealgequiv}) we have
\[
\sum_{j=1}^{2g+1}\partial_{u_j}q_{g+1}(\tilde{u_1},\tilde{u_2},\vu)+
\partial_{\tilde{u_1}}q_{g+1}(\tilde{u_1},\tilde{u_2},\vu)
+\partial_{\tilde{u_2}}q_{g+1}(\tilde{u_1},\tilde{u_2},\vu)=
\dfrac{1}{\pi i }\int_{\li_g\cup\tilde{I}}
\dfrac{\rho^{\prime}(z)}{Y^+(z) }dz.
\]
The limit $\tilde{u_1}\ra\tilde{u_2}$ of the left hand side of the 
above identity has been obtained in (\ref{Philimit}). Here we 
 derive the limit of the right hand side.
Let us define $\tilde{u_1}=\lb+\sqrt\d,\;\tilde{u_2}=\lb-\sqrt\d$.
Then
\begin{equation}
\nonumber
\begin{split}
\lim_{\d\ra 0}\dfrac{1}{\pi
i}\int_{\li_g\cup\tilde{I}}
\dfrac{2\rho^{\prime}(z)}{Y^+(z)}dz&=
\dfrac{1}{\pi
i}\int_{\li_g}
\dfrac{2\rho^{\prime}(z)}{y^+(z)(z-\lb)}dz+
\lim_{\d\ra 0}\dfrac{1}{\pi
i}\int_{\lb-\sqrt\d}^{\lb+\sqrt\d}
\dfrac{2\rho^{\prime}(z)}{Y^+(z)}dz\\
=&\dfrac{1}{\pi
i}\int_{\li_g}
\dfrac{2\rho^{\prime}(z)}{y^+(z)(z-\lb)}dz-
\dfrac{2\rho^{\prime}(\lb)}{y(\lb)}.
\end{split}
\end{equation}
Combining the previous three relations we obtain
\begin{equation}
\label{Phigap}
2\Phi^g(\lb;\vu)=\dfrac{1}{\pi
i}\int_{\li_g}
\dfrac{2\rho^{\prime}(z)}{y^+(z)(z-\lb)}dz-\dfrac{2\rho^{\prime}(\lb)}{y(\lb)},\;\;
\lb\in(u_{2j+1},u_{2j}),\;j=0,\dots,g,\;u_0=1.
\end{equation}

Combining (\ref{GH}) and  (\ref{Phigap})
 the variational condition (\ref{lax0}) can be easily obtained.
\hfill $\square$
\begin{rem}
We observe that the function $\Phi^g(\lb;\vu)-6t\d_{g0}$ must be 
nonzero at the branch points. This is obvious for g=0. Indeed
$\Phi^0(u;u)-6t=f^{\prime}(u)-6t=1/u_x(x,t)$.
For $g>0$ the derivatives $\partial_x u_i(x,t)$  are given by the
expression  \cite{G1}
\[
\partial_x u_i(x,t)=\dfrac{P_0^g(u_i)} { 2\prod_{\substack{j=1\\j\neq 
i}}^{2g+1}(u_i-u_j)\Phi^g(u_i;\vu)},\quad i=1,\dots,2g+1,\;g>0,
\]
where $P_0^g$ has been defined in (\ref{b6}). From the above
expression
it is obvious that the $\partial_x u_i(x,t)$'s are nonsingular if
$\Phi^g(u_i;\vu)\neq 0$, $i=1,\dots,2g+1$.
\end{rem}
\subsubsection{Phase transitions}
We derive the equations which determine a change of  genus of
the solution of the hodograph transformation  (\ref{ch})
 or the set of algebraic
equations (\ref{moment}) and (\ref{normalization}).
We observe that a transition necessarily occurs when one of the
two conditions (\ref{RHg11}) or (\ref{lax0}) fail to be satisfied.
If the conditions (\ref{RHg11}) fail to be satisfied at 
some point $v$ in the bands $(u_{2j},u_{2j-1}),\;j=1,\dots,g+1$,
it follows that 
\begin{equation}
\label{trail}
\begin{split}
&\int_{v}^{u_{2j+1}}
y(\xi)(\Phi^g(\xi;\vu)-6t\e_{g0})d\xi=0,\\
&\Phi^g(v;\vu)-6t\e_{g0}=0\,\quad
v\in(u_{2j},u_{2j-1}),\;j=1,\dots,g+1.
\end{split}
\end{equation}	
Indeed $v$ must be a stationary point of the integral $\int_{\lb}^{u_{2j+1}}
y(\xi)(\Phi^g(\xi;\vu)-6t\e_{g0})d\xi$.
When we can solve the above system together with the moment conditions
(\ref{moment}) and the normalization conditions (\ref{normalization})
 for some $t>0$, we obtain a point $x(t)$,
$v(t)$ and  $u_1(t)>u_2(t)>\dots>u_{2g+1}(t)$  of the boundary 
between the $g$-phase solution and the $(g+1)-$phase solution.
 
\noindent
In the same way if the conditions (\ref{lax0}) fail to be satisfied at 
some point $v$ in the gaps $(u_{2j},u_{2j+1}),\;j\!=\!0,\dots,g$,
it follows that 
\begin{equation}
\label{lead}
\begin{split}
&\Phi^g(v;\vu)-6t\e_{g0}=0\\
&\partial_v \Phi^g(v;\vu)=0.
\end{split}
\end{equation}
The above system describes the point of phase transition when a new
band is opening.

\noindent
Systems (\ref{trail}) and (\ref{lead}) 
have been obtained in \cite{G1}
studying directly the hodograph transformation (\ref{ch}).
%%%%%%%%%%%%%%%%%%%%%%%%%%%%%%%%%%%%%%%%%%%%%%%%%%%%%%%%%%%%%%%%%%%%%%%%%%%%%%%%%%%%%%%%%%%%%%%%%%%%%%%%%%%%%%%%%%%%%%%%%%%%%%%%%%%%%%%%%%%%%%%%%%%%%%%%%%%%%%%%%%%%%%%%%%%%%%%%%%%%%%%%%%%%%%%%%%%%%%%%%%%%%%%%%%%%%%%%%%%%%%%%%%%%%%%%%%%%%%%%%%%%%%%%%%%%%%%%%%%%%%%%%%%%%%%%%%%%%%%%%%%%%%%%%%%%%%%%%%%%%%%%%%%%%%%%%%%%%%%%%%%%%%%%%%%%%%%%%%%%%%%%%%%%%%%%%%%%%%%%%%%%%%%%%%%%%%%%%%%%%%%%%%%%%
\subsection{Lax-Levermore-Venakides functional}

Following the steps in \cite{DVZ} we  construct the maximizer of
the Lax-Levermore-Venakides functional \cite{LL},\cite{V}.
Let us make the change of variable $\lb=1-\eta^2$, taking the upper
complex half plane onto $C\backslash (-\infty,1]$.  The function
$\g(\lb)$ transforms to $\f(\eta)=\g(\lb)$. The function $\f$ is
analytic
off the real $\eta$ axis. We extend the definition of $\f$
onto the lower complex $\eta$-plane by the relation $\f(-\eta)=-\f(\eta)$.
The RH problem for the function $\f$ is
$\f_+-\f_-=\g_++\g_-$ and $\f_++\f_-=-(\g_+-\g_-)sign\eta$.
On the real axis we define the function $\psi^*(\eta)$ to be equal to
zero outside the interval $[-1,1]$, while on the interval $(-1,1)$
\begin{equation}
\label{lax}
\psi^*(\eta)=\partial_{\eta}[\a(\lb(\eta))-\rho(\lb(\eta))+
\dfrac{1}{2}(\f_+-\f_-)].
\end{equation}
The function $\psi^*(\eta)$ is the unique maximizer  
of the functional \cite{V},\cite{DVZ}
\[
Q(\psi)=\dfrac{1}{\pi}[(2a,\psi)+(L\psi,\psi)],\quad \psi\in
L^1([0,1]),\;\psi\leq 0,
\]
where
\[
L\psi(\eta)=\dfrac{1}{\pi}\int_0^1\ln\left|\dfrac{\eta-\mu}{\eta+\mu}\right|\psi(\mu)d\mu
\]
\[
a(\eta,x,t)=4t\eta^3-x\eta-6t\eta+\dfrac{1}{2}\int_{1-\eta^2}^1
\dfrac{f(\xi)}{\sqrt{\xi+\eta^2-1}}.
\]

The maximization problem is attacked analytically by solving the
variational conditions
\begin{equation}
\label{lax1}
\begin{split}
L\psi(\eta)+a(\eta,x,t)&=0,\quad \mbox{when}\;\psi(\eta,x,t)<0\\
L\psi(\eta)+a(\eta,x,t)&>0,\quad \mbox{when}\;\psi(\eta,x,t)=0.
\end{split}
\end{equation}
Namely if $\psi$ satisfies (\ref{lax1}), then $\psi=\psi^*$ \cite{LL}.
In \cite{DVZ} it is shown that the variational conditions 
(\ref{lax1}) transform to (\ref{RHg1}) and (\ref{RHg2}). The benefit
of the variational formulation is that, due to the convexity of the
maximization
problem, the uniqueness of $\psi^*$ and hence of $\g$ is guaranteed.
Furthermore for each fixed $x$ and $t$ the support of the maximizer
in uniquely defined. An important consequence of this result is the following. 
\begin{theo}
For almost all $x$ and $t\geq 0$ the solution 
$u_1(x,t)>u_2(x,t)>\dots>u_{2g+1}(x,t)$ of the hodograph transformation (\ref{ch})
\[
x=-v_i(\vu)+w_i(\vu), \quad i=1,\dots,2g+1,\;g\geq 0,
\]
which satisfies  the constraints (\ref{RHg1}) and (\ref{RHg2}) or,
equivalently, (\ref{RHg11}) and (\ref{lax0}), 
exists for some $g\geq 0$ and it is unique.
\end{theo}
As a final result we write the Lax-Levermore-Venakides maximizer as 
the solution of a linear overdetermined system of
Euler-Poisson-Darboux type.
\begin{theo}
For $g>0$ the maximizer $\psi^*(\eta)$ in (\ref{lax}) can be written in the form
\begin{equation}
\nonumber
\begin{split}
\nonumber
\psi^*(\eta;\b_1,\b_2,\dots,\b_{2g+1})&=0, \quad \eta\in(\b_1,\b_2)\cup(\b_3,\b_3)\cup\dots\cup(\b_{2g+1},1)\\
\psi^*(\eta;\b_1,\b_2,\dots,\b_{2g+1})&=-2\eta\,
\Phi^g(1-\eta^2;1-\b_1^2,1-\b_2^2,\dots,1-\b_{2g+1}^2)y(1-\eta^2)\;\;\\
&\mbox{when}\;
\;\eta\in[0,\b_1]
\cup_{j=1}^g[\b_{2j},\b_{2j+1}],\;
\end{split}
\end{equation}
\[
\beta_k=\sqrt{1-u_k},\quad k=1,\dots,2g+1,
\]
\[
y(1-\eta^2)=\prod_{j=1}^{2g+1}(1-\eta^2-u_j)^{\frac{1}{2}}
\]
and the function $\Phi^g$ has been defined in (\ref{Phi}). 
\end{theo}
We observe that the condition  (\ref{lax0})
implies $ \psi^*(\eta;\b_1,\b_2,\dots,\b_{2g+1})<0$ for $\eta\in(0,\b_1)
\cup_{j=1}^g(\b_{2j},\b_{2j+1})$ and vice-versa.
The proof of the theorem follows directly from (\ref{lax0}) and
(\ref{lax}).
\begin{rem}
All the information on the initial data of the Lax-Levermore
 maximizer is contained in the the function $\Phi^g(\lb;\vu)$.

\noindent
In \cite{G1},\cite{G}  an upper bound to the genus of the solution of the Whitham
equations was provided. We selected initial data such that the maximum
number of real zeros of the function $\Phi^g(\lb;\vu)-6t\d_{g0}$
  is $2N-g$ for $0\leq g\leq N$
where $N$ is the supposed upper bound of the genus. 
Then we showed  that, in such a situation, 
phase transitions from solutions of genus $g\leq N$ to  solutions of
genus  $g>N$ do not occur. 
The proof of the theorem is obtained  only studying the Whitham equations
and the hodograph transformation (\ref{ch}).

\noindent
In this new representation of the Lax-Levermore-Venakides maximizer,
the above result has an easy interpretation. Namely it gives
an upper bound to number of intervals   where the function
$y(\lb)(\Phi^g(\lb;\vu)-6t\d_{g0})$ is positive for $0\leq g\leq N$. 
\end{rem}

\section{Conclusion}
In this paper
 we have studied the Cauchy problem for the KdV equation with small
dispersion and with monotonically increasing initial data.
We have used   the formulation of  the 
 Cauchy problem for KdV as a Riemann-Hilbert (RH) problem  \cite{S}.
 The small $\e$-asymptotics
is obtained using 
 the steepest descent  method for oscillatory Riemann-Hilbert problems
introduced in \cite{DZ}.  Analyticity of the initial data is essential 
for this  step.

\noindent
This procedure leads to a scalar RH problem for a function $\g$
and a  set of algebraic equations constrained by algebraic
inequalities.

The scalar RH problem for the function $\g$ 
 is well defined even for smooth initial
data.

\noindent
In this paper
 we have shown that, for smooth monotonically increasing initial data
bounded at infinity, the set of algebraic equations obtained
through the Deift, Venakides and Zhou approach can be expressed as
  the solution of  a linear overdetermined systems of equations of
Euler-Poisson-Darboux type.
 We have  also shown that this set of algebraic equation is equivalent
to the set of algebraic equations defined by 
the hodograph transformation  (\ref{ch}). The uniqueness of the Lax-Levermore
maximization problems guarantees the uniqueness of the solution of
the Cauchy problem for  the Whitham equations.
As a final result we have  shown that the Lax-Levermore maximizer can 
also be expressed as a solution of a linear overdetermined system
of Euler-Poisson-Darboux type.

\noindent 
We believe that the 
above results can be extended to bump-like  initial data.

\vskip 0.5cm
\noindent 
{\bf Acknowledgments.} I am very grateful to   
 Percy Deift who  explained me the  steepest descent  method
 for oscillatory Riemann-Hilbert problems.

\noindent
I wish to thank  Yang  Chen for pointing out the proof of
theorem~\ref{equiv}.

\noindent
I am indebted to Boris Dubrovin for many discussions during the 
preparation of this  manuscript.

\noindent
This worked has been supported with an EC grant No HPMF-CT-1999-00263.

\end{document}